\documentclass{jfm}
\usepackage{todonotes}
\usepackage{graphicx}
\usepackage{epstopdf, epsfig}
\usepackage{amsmath,amsfonts,amssymb}
\usepackage{float}
\usepackage{gensymb}
\usepackage{mathrsfs}
\usepackage{subfigure}
\usepackage{subfig}
\usepackage{url}
\usepackage{natbib}
\usepackage{todonotes}
\usepackage{soul}

\newcommand{\Me}{\text{Me}}
\newcommand{\otol}{$(1,2,1)^{\text{L}}$}
\newcommand{\otoh}{$(1,2,1)^{\text{H}}$}
\newcommand{\alp}{\alpha_k}
\newcommand{\ups}{\Upsilon_k}
\newcommand{\balp}{\overline{\alpha}_k}
\newcommand{\bups}{\overline{\Upsilon}_k}

\shorttitle{Emergence of superwalking droplets}
\shortauthor{R. N. Valani, J. Dring, T. P. Simula and A. C. Slim}

\title{Emergence of superwalking droplets}

\author{Rahil N. Valani\aff{1}
  \corresp{\email{rahil.valani@monash.edu}},
  Jack Dring\aff{2},
  Tapio P. Simula\aff{3}
  and Anja C. Slim\aff{2,4}}

\affiliation{\aff{1}School of Physics and Astronomy, Monash University, Victoria 3800, Australia
\aff{2}School of Mathematics, Monash University, Victoria 3800, Australia
\aff{3}Optical Sciences Centre, Swinburne University of Technology, Melbourne 3122, Australia
\aff{4}School of Earth, Atmosphere and Environment, Monash University, Victoria 3800, Australia}

\begin{document}

\maketitle

\begin{abstract}
A new class of self-propelled droplets, coined superwalkers, has been shown to emerge when a bath of silicone oil is vibrated simultaneously at a given frequency and its subharmonic tone with a relative phase difference between them \citep{superwalker}. To understand the emergence of superwalking droplets,  we explore their vertical and horizontal dynamics by extending previously established theoretical models for walkers driven by a single frequency to superwalkers driven by two frequencies. Here we show that driving the bath at two frequencies with an appropriate phase difference raises every second peak and lowers the intermediate peaks in the vertical periodic motion of the fluid surface. This allows large droplets that could otherwise not walk to leap over the intermediate peaks, resulting in superwalking droplets whose vertical dynamics is qualitatively similar to normal walkers. We find that the droplet's vertical and horizontal dynamics are strongly influenced by the relative height difference between successive peaks of the bath motion, a parameter that is controlled by the phase difference. Comparison of our simulated superwalkers with the experiments of \citet{superwalker} shows good agreement for small- to moderate-sized superwalkers.
\end{abstract}


\section{Introduction}
On vertically vibrating a bath of silicone oil at frequency $f$, a droplet of the same oil can be made to bounce indefinitely on the free surface of the liquid \citep{Walker1978,Couder2005}. As the amplitude of the forcing is increased, the bouncing droplet destabilises and transitions to a steady walking state \citep{Couder2005WalkingDroplets}. The walking droplet, also called a `walker', emerges just below the Faraday instability threshold \citep{Faraday1831a}, above which the whole surface becomes unstable to standing Faraday waves oscillating at the subharmonic frequency $f/2$. On each bounce, the walker generates a localised damped Faraday wave on the fluid surface. It then interacts with these waves on subsequent bounces, giving rise to a self-propelled wave-droplet entity. Intriguingly, such walkers have been shown to mimic several peculiar behaviours that were previously thought to be exclusive to the quantum world. These include orbital quantisation in rotating frames \citep{Fort17515,harris_bush_2014,Oza2014} and harmonic potentials \citep{Perrard2014b,Perrard2014a,labousse2016}, Zeeman splitting in rotating frames \citep{Zeeman,spinstates2018}, wave-like statistical behaviour in confined geometries \citep{PhysRevE.88.011001,Giletconfined2016,Saenz2017,Cristea,durey_milewski_wang_2020} as well as in an open system \citep{Friedal} and tunnelling across submerged barriers \citep{Eddi2009,tunnelingnachbin,tunneling2020}. They have also been predicted to show anomalous two-droplet correlations \citep{ValaniHOM,correlationnachbin}. A detailed review of hydrodynamic quantum analogues of walking droplets is provided by \cite{Bush2015} and \cite{Bush2018}. 

 Recently, a new class of walking droplets, coined \emph{superwalkers}, has been observed \citep{superwalker}. These emerge when the bath is driven simultaneously at two frequencies, $f$ and $f/2$, with a relative phase difference $\Delta\phi$. For a commonly studied system with silicone oil of $20\,\text{cSt}$ viscosity, single frequency driving at $f=80\,\text{Hz}$ produces walkers with radii between $0.3\,\text{mm}$ and $0.5\,\text{mm}$, and walking speeds up to $15\,\text{mm}/\text{s}$, with speed typically increasing with size \citep{Molacek2013DropsTheory,Wind-Willassen2013ExoticDroplets}. In the same system with two frequency driving at $f=80$\,Hz and $f/2=40$\,Hz, superwalkers can be significantly larger than walkers with radii up to $1.4\,\text{mm}$ and walking speeds up to $50\,\text{mm}/\text{s}$ \citep{superwalker}.  Intriguingly, the walking speed and the vertical dynamics of superwalkers are strongly dependent on the phase difference $\Delta\phi$, with peak superwalking speed occurring near $\Delta\phi=140^{\circ}$, while near $\Delta\phi=45^{\circ}$ they only bounce or may even coalesce.  For a fixed phase difference, smaller superwalkers typically behave very similarly to walkers, with their speed increasing with their size and impacting the surface once every two up-and-down cycles of the bath. Conversely, the speed of larger superwalkers decreases with their size. These large superwalkers appear to impact the bath twice every two up-and-down cycles of the bath and have prolonged contact with the bath, with the largest ones hardly lifting from the surface. Using sophisticated numerical simulations, \citet{galeano-rios_milewski_vanden-broeck_2019} were able to replicate superwalking behaviour for a single droplet of moderate radius $R=0.68$\,mm, and reported a good match in the superwalking speed between their simulation and the experiments of \cite{superwalker}. Although these two studies describe the characteristics of superwalkers, an understanding of the mechanism that enables superwalking is still lacking. In this paper, our aim is to understand this underlying mechanism by adapting the theoretical models used for walkers driven with a single frequency, to superwalkers driven with two frequencies.

Over the years, many theoretical models have been developed to describe the dynamics of a walker. These range from phenomenological stroboscopic models that only capture the horizontal dynamics to sophisticated models that resolve the vertical and horizontal dynamics and the detailed evolution of the surface waves created by the walker. Intermediate complexity models that resolve the vertical and horizontal dynamics but assume a predetermined form for the standing wave generated by the droplet on each impact have been widely used. In this latter category, \citet{molacek_bush_2013} modelled the vertical bouncing dynamics of the droplet using a linear spring model, inspired by the investigations of \citet{Chaoticbouncing,gilet_bush_2009} on droplets bouncing on a soap film. They also developed a nonlinear logarithmic spring model, although it is not clear whether this model is more accurate and hence the linear spring model is often used for simplicity \citep{couchman_turton_bush_2019}. \citet{Molacek2013DropsTheory} and \cite{couchman_turton_bush_2019} coupled these vertical spring models with horizontal dynamics comprising a propulsive force from the impact and a lumped drag force consisting of aerodynamic drag and momentum drag during contact.  The propulsive force is the horizontal component of the normal force, which arises because the small-amplitude waves incline the free surface. The waves are modelled by a linear superposition of predetermined standing waves generated by the droplet on each impact.  \citet{milewski_galeano-rios_nachbin_bush_2015} further refined the modelling by coupling the vertical and horizontal dynamics models of \citet{molacek_bush_2013,Molacek2013DropsTheory} to a quasi-potential, weakly viscous wave model for wave generation and evolution. \citet{galeano-rios_milewski_vanden-broeck_2017} developed a more complete model for the vertical dynamics of the droplet by imposing a kinematic match between the motion of the free surface and that of the impacting droplet, modelled as a solid sphere. Combining this vertical dynamics model with the free surface evolution of \citet{milewski_galeano-rios_nachbin_bush_2015}, \citet{galeano-rios_milewski_vanden-broeck_2019} were able to obtain an accurate model for walking droplets free of any impact parametrisation. Such models give an accurate description of the system at the time scale of a single bounce but they become inefficient in calculating a droplet's horizontal trajectory over long times. Hence, stroboscopic models that average over the walker's periodic vertical motion but capture the horizontal motion have been developed to investigate the horizontal dynamics of walkers over long time scales. Many of these stroboscopic models use a predetermined form of the standing wave field \citep{Protiere2006,Oza2013,PhysRevFluids.3.013604,PhysRevFluids.2.053601,couchman_turton_bush_2019} while the model of \citet{durey_milewski_2017} uses a more sophisticated wave model to accurately capture the droplet's wave field. A comprehensive review of the different walker models is given by \citet{Turton2018}.
 
In this study, we couple the vertical and horizontal dynamics models of \cite{molacek_bush_2013,Molacek2013DropsTheory} along with a new model for the wave field of a superwalker to understand and rationalise superwalking. In \S\:\ref{NM}, we provide a summary of the theoretical model, explore the wave field for two-frequency driving and describe the nomenclature we use to describe the bouncing modes. In \S\:\ref{ES}, we show that adding the second driving frequency with an appropriate phase difference raises every second peak and lowers the intermediate peaks of the bath's motion, and that this allows larger droplets to leap over the intermediate peaks, thereby enabling superwalking. We also show the importance of the phase difference $\Delta\phi$ in the emergence of superwalking and compare the results from simulations with the experiments of \citet{superwalker}. We discuss and conclude the study in \S\:\ref{section: DC}.
 
\section{Theoretical model}\label{NM}
\begin{figure}
\centering
\includegraphics[width=\columnwidth]{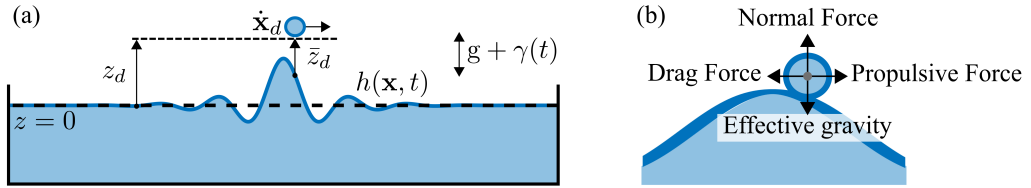}
\caption{(a) Schematic of the system consisting of a bath of liquid vibrated vertically with acceleration $\gamma(t)$ and a droplet of the same liquid walking horizontally at velocity $\dot{\mathbf{x}}_d$ and located at vertical position $z_d$ relative to the free surface of the liquid at rest. Panel (b) shows the vertical and horizontal forces acting on the droplet in the comoving frame of the bath. In the vertical direction, the droplet experiences an effective gravity, $-m[\text{g}+\gamma(t)]$, and an upward normal force,  $F_N(t)$, during contact. In the horizontal direction the droplet experiences a propulsive force, $-F_N(t) \nabla h(\mathbf{x}_d,t)$, during contact due to the slope of the wave field and a lumped drag force composed of momentum loss during contact,  $-D_{mom}\dot{\mathbf{x}}_d$ and air drag, $-D_{air}\dot{\mathbf{x}}_d$.}
\label{fig: schematic}
\end{figure}

Consider a droplet of mass $m$ and radius $R$ bouncing on a bath of liquid of density $\rho$, kinematic viscosity $\nu$ and surface tension $\sigma$. The bath is vibrating vertically with acceleration $\gamma(t)=\Gamma_f\text{g}\sin(\Omega t)+\Gamma_{f/2}\text{g}\sin(\Omega t/2+\Delta\phi)$. Here $\Omega=2\pi f$ is the angular frequency, $\Gamma_f$ and $\Gamma_{f/2}$ are the acceleration amplitudes of the two frequencies relative to gravity g, and $\Delta\phi$ is the relative phase difference between them. This configuration is shown schematically in figure~\ref{fig: schematic}. We describe our system in the oscillating frame of the bath by horizontal coordinates $\mathbf{x} = (x,y)$ and vertical coordinate $z$, with $z=0$ chosen to coincide with the undeformed surface of the bath. In this frame, the centre of mass of the droplet is located at a horizontal position $\mathbf{x}_d$ and the south pole of the droplet at a vertical position $z_d$ such that $z_d=0$ would represent initiation of contact with the undeformed surface of the bath. The free surface elevation of the liquid filling the bath is at $z=h(\mathbf{x},t)$.

\subsection{Vertical dynamics}\label{vd}
We simulate the vertical droplet dynamics using the linear spring model of \citet{molacek_bush_2013} adapted for two-frequency driving. Using this model, the vertical equation of motion of the droplet in the comoving frame of the bath is given by
\begin{equation}\label{eq: vertical}
    m\ddot{z}_d=-m[\text{g}+\gamma(t)]+{F_N}(t),
\end{equation}
where the first term on the right hand side is the effective gravitational force on the droplet in the bath's frame of reference. The second term on the right hand side is the normal force imparted to the droplet during contact with the liquid surface. This normal force is calculated by modelling interaction with the bath as a linear spring and damper according to, 
\begin{equation}
{F_N}(t)=H(-\bar{z}_d) \,\max\left(-k_s\bar{z}_d-b\dot{\bar{z}}_d,0\right).
\label{springdamp}
\end{equation}
Here $H$ is the Heaviside step function and $\bar{z}_d=z_d-h(\mathbf{x}_d,t)$ is the height of the droplet's south pole above the free surface of the bath (from here on simply referred to as the height of the droplet). The maximum condition in \eqref{springdamp} ensures a non-negative reaction force on the droplet during contact. The constants $k_s$ and $b$ are the spring constant and damping force coefficient, respectively. We discuss appropriate values for these constants in \S\:\ref{nmpv}.

\subsection{Horizontal dynamics}
To describe the horizontal dynamics of the walking droplet, we use the model of \citet{Molacek2013DropsTheory} for which the horizontal equation of motion is 
\begin{equation}\label{eq: hor}
    m\ddot{\mathbf{x}}_d=-\left[D_{mom}(t)+D_{air}\right]\dot{\mathbf{x}}_d-{F_N}(t) \boldsymbol{\nabla} h(\mathbf{x}_d,t).
\end{equation}
 The term in parentheses on the right hand side is the total instantaneous drag force, composed of momentum loss during contact, $D_{mom}(t)=C\sqrt{\rho R/\sigma}{F_N}(t)$, and an air drag of the form $D_{air}=6\pi R \mu_a$. Here $\mu_a$ is the dynamic viscosity of air and $C$ is the contact drag coefficient, an adjustable parameter. We discuss appropriate values for this coefficient in \S\:\ref{nmpv}. The final term on the right hand side is the horizontal component of the contact force arising from the small slope, $|\boldsymbol{\nabla} h(\mathbf{x}_d,t)|<<1$, of the underlying wave field during contact.
 
\subsection{Wave field}
The free surface $z=h(\mathbf{x},t)$ is calculated as the linear superposition of all the individual waves generated by the droplet on its previous bounces,
\begin{equation*}
\textstyle h(\mathbf{x},t)=\sum_{n} h_n(\mathbf{x},t)\, ,
\end{equation*}
where $h_n(\mathbf{x},t)$ is the wave field generated by the $n$th bounce at location $\mathbf{x}_n$ and time $t_n$. 

The individual waves generated by the droplet on each bounce are localised, damped standing Faraday waves. Various different models of the waveform have been developed to describe a single impact of a walker \citep{eddi_sultan_moukhtar_fort_rossi_couder_2011,Molacek2013DropsTheory,milewski_galeano-rios_nachbin_bush_2015,tadrist_shim_gilet_schlagheck_2018}. One of the most commonly used wave model is that of \citet{Molacek2013DropsTheory}, given by
\begin{equation} \label{molacek wave}
h^{(M)}_n(\mathbf{x},t)=\frac{A^{(M)}}{\sqrt{t-t_n}}\cos\left(\frac{\Omega t}{2}\right)J_0(k_F|\mathbf{x}-\mathbf{x}_n|)\,\exp{\left[-\frac{t-t_n}{T_F \Me^{(M)}}\right]},
\end{equation}
where $k_F$ is the Faraday wavenumber, $T_F=4\pi/\Omega$ is the Faraday period, and $\Me^{(M)}=T_d/T_F(1-\Gamma_f/\Gamma_F)$ is the memory parameter that determines the proximity to the Faraday threshold.  In this latter expression,  $T_d=1/\nu_e k_F^2$ is the time constant for wave decay, $\nu_e$ is the effective kinematic viscosity and $\Gamma_F$ is the dimensionless acceleration amplitude at the Faraday threshold for single-frequency driving at frequency $f$. This model describes a wave with the shape of a Bessel function of the first kind, $J_0$, that oscillates at the subharmonic frequency $f/2$ and decays exponentially in time with a decay constant inversely proportional to the memory. The location and instant of the droplet's impact are approximated respectively by,
\begin{equation}\label{wa_pos}
    \mathbf{x}_n=\int_{t_n^i}^{t_n^c}\mathbf{x}_d(t')F_N(t')\,\text{d}t'\Big{/}\int_{t_n^i}^{t_n^c}F_N(t')\,\text{d}t'\:\:\text{and}\:\:t_n=\int_{t_n^i}^{t_n^c}t'F_N(t')\,\text{d}t'\Big{/}\int_{t_n^i}^{t_n^c}F_N(t')\,\text{d}t',
\end{equation}
where $t_n^i$ and $t_n^c$ are the time of initiation and completion of the $n$th impact. The equation for the wave amplitude coefficient $A^{(M)}$ is
\begin{equation*}
    A^{(M)}=\sqrt{\frac{2\nu_e}{\pi}}\frac{k_F^3}{3\sigma k_F^ 2 + \rho \text{g}}\int_{t_n^i}^{t_n^c}\sin\left(\frac{\Omega t'}{2}\right)F_N(t') \,\text{d}t'.
\end{equation*}

A detailed theoretical study of the wave field generated by a single bounce of a walker was undertaken by \citet{tadrist_shim_gilet_schlagheck_2018}. They derived the following improved waveform for the wave generated by an instantaneous impact of a walker of force strength $F_0$ at location $\mathbf{x}_n$ and time $t_n$,
\begin{equation}\label{tadrist_wave}
h^{(T)}_n(\mathbf{x},t)=\frac{A_0^{(T)}}{\sqrt{t-t_n}}\cos\left(\frac{\Omega t}{2}+\theta^{+}_F\right)J_0(k_F|\mathbf{x}-\mathbf{x}_n|)\,\exp{\left[-\frac{(t-t_n)}{T_F \Me^{(T)}}-\frac{T_F|\mathbf{x}-\mathbf{x}_n|^2}{8\pi D(t-t_n)}\right]}.   
\end{equation}

In this expression, the memory parameter is given by $\Me^{(T)}=-1/2\pi\delta^{+}_F$ with $\delta^{+}_F$ the dimensionless decay rate of the longest-lived Faraday wave. This improved form of the wave field has two new additions: (i) the phase shift $\theta^{+}_F$ of the Faraday waves relative to the driving signal and (ii) an exponential spatial decay with diffusive spreading (with a diffusion coefficient $D$). Note that similar additions can also be obtained following the derivation of \citet{Molacek2013DropsTheory} by including higher-order terms in their decay rate expansions. The amplitude coefficient $A_0^{(T)}$ takes the form, 
\begin{equation*}
    A_0^{(T)}=\sqrt{\frac{2\pi}{\Omega^5 D}}\frac{2k_F^2}{\pi\rho}B^{+}_F(t_n)F_0,
\end{equation*}
where $B^{+}_F(t_n)$ is a function that prescribes the amplitude based on the instant of impact $t_n$ and is given by
\begin{equation}\label{bffunc}
B^{+}_F(t_n) =\frac{-2\cos(\Omega  t_n/2+\theta_{F}^{-})}{(\delta^{+}_F-\delta^{-}_F)[\cos(\Omega t_n+\theta^{+}_{F}+\theta^{-}_{F})+\cos(\theta^{+}_{F}-\theta^{-}_{F})]-2\sin(\theta^{+}_{F}-\theta^{-}_{F})}, 
\end{equation}
where $\theta^{-}_{F}$ and $\delta^{-}_{F}$ are the phase shift and decay rate of a companion, short-lived Faraday wave.  The reader is referred to \citet{tadrist_shim_gilet_schlagheck_2018} for further details on these parameters. To extent this model to a finite contact time, we follow the suggestion in \citet{tadrist_shim_gilet_schlagheck_2018} of using Duhamel's principle and the approach used in Appendix A.4 of \citet{Molacek2013DropsTheory}, and integrate the impulse response with a time varying contact force $F_N(t)$. This results in replacing the amplitude coefficient $A_0^{(T)}$ in \eqref{tadrist_wave} by
\begin{equation}\label{tad_amp}
    A^{(T)}=\sqrt{\frac{2\pi}{\Omega^3 D}}\frac{k_F^2}{\pi \rho}\int_{t_n^i}^{t_n^c}B^{+}_F(t') F_N(t') \,\text{d}t',
\end{equation}
 and replacing the initial impact location $\mathbf{x}_n$ and time $t_n$ with their respective weighted averages as defined in \eqref{wa_pos}. Moreover, in \eqref{tad_amp}, the amplitude prescribing function $B_F^{+}(t')$ is identical to the one presented in \eqref{bffunc}.

To model the wave field generated by a superwalker with two-frequency driving, we use a similar approach to that of \citet{tadrist_shim_gilet_schlagheck_2018}.  A detailed derivation is presented in Appendix \ref{der}. The final form of the wave field for the case of the bath being driven at $f=80$\,Hz and $f/2=40$\,Hz is
\begin{align}\label{sw_wave}
h_n^{(\text{SW})}(\mathbf{x},t)= A_{40}\frac{\cos(\Omega t/2+\theta_{F40}^{+})}{\sqrt{t-t_n}} \,\text{J}_0(k_{F40}|\mathbf{x}-{\mathbf{x}}_n|)\,\exp{\left[ -\frac{t-{t}_n}{T_F \text{Me}_{40}} - \frac{T_F |\mathbf{x}-{\mathbf{x}}_n|^2}{8 \pi D_{40} (t-{t}_n)} \right]}& \nonumber\\
+ A_{20}\frac{\cos(\Omega t /4 + \theta_{F20}^{+})}{\sqrt{t-t_n}} \,\text{J}_0(k_{F20}|\mathbf{x}-{\mathbf{x}}_n|)\,\exp{\left[ -\frac{t-{t}_n}{T_F\text{Me}_{20}} - \frac{T_F |\mathbf{x}-{\mathbf{x}}_n|^2}{8 \pi D_{20} (t-t_n)} \right]}&,
\end{align}
where the impact location $\mathbf{x}_n$ and the time of impact $t_n$ are calculated using \eqref{wa_pos}, and the amplitude coefficients are given by
\begin{align}\label{amp_sw}
    A_{40}&=\sqrt{\frac{2\pi}{\Omega^3 D_{40}}}\frac{k_{F40}^2}{\pi \rho}\int_{t_n^i}^{t_n^c}B^{+}_{F40}(t') F_N(t') \,\text{d}t',\nonumber\\
    A_{20}&=\sqrt{\frac{2\pi}{\Omega^3 D_{20}}}\frac{k_{F20}^2}{\pi \rho}\int_{t_n^i}^{t_n^c}B^{+}_{F20}(t') F_N(t') \,\text{d}t'.
\end{align}
The interpretation of \eqref{sw_wave} is that a droplet bouncing under the prescribed two-frequency driving excites two dominant waves at wavenumbers $k_{F40}$ and $k_{F20}$, corresponding to  frequencies of $40$\,Hz and $20$\,Hz.  These waves decay in time at rates $\text{Re}(\delta^{+}_{F40})$ and $\text{Re}(\delta^{+}_{F20})$, which determine the corresponding memory parameters $\Me_{40}=-1/2\pi\text{Re}(\delta^{+}_{F40})$ and $\Me_{20}=-1/2\pi\text{Re}(\delta^{+}_{F20})$. Here $\text{Re}(\cdot)$ denotes the real part of the complex argument. The waves also spread diffusively with diffusion coefficients $D_{40}$ and $D_{20}$, and have phase shifts $\theta_{F40}^{+}$ and $\theta_{F20}^{+}$ with respect to the driving signal. We refer the reader to Appendix \ref{der} for explicit equations for these parameters. 

 \begin{figure}
\centering
\includegraphics[width=\columnwidth]{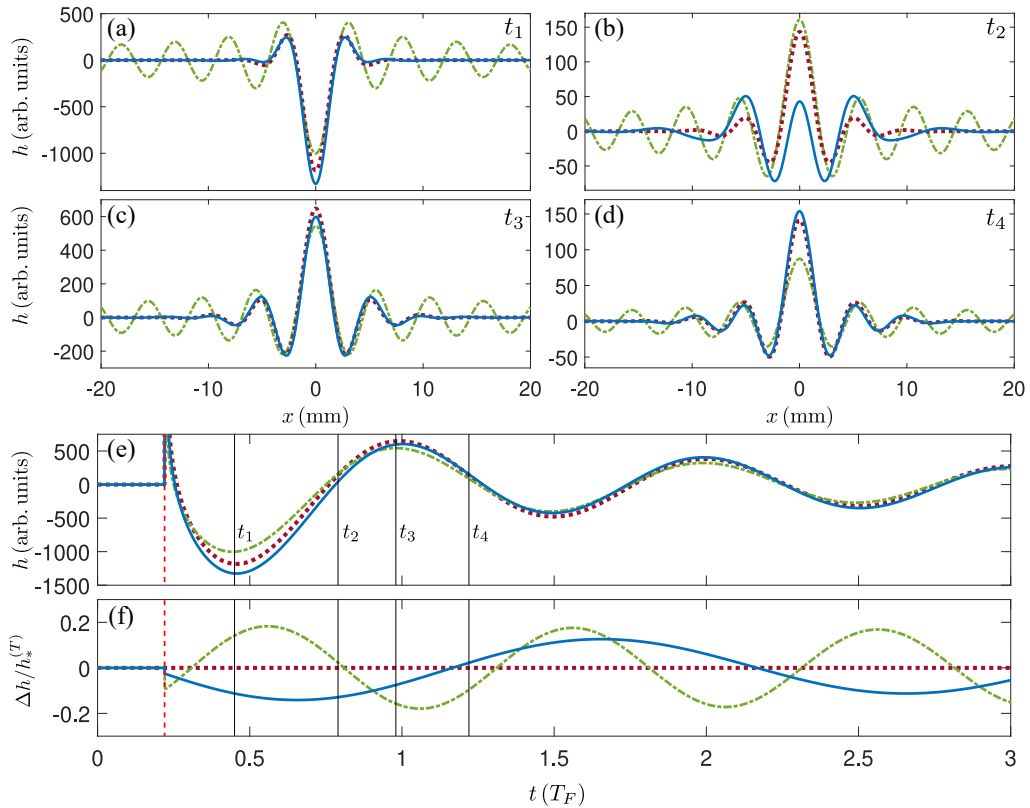}
\caption{Comparison of the wave fields generated by an instantaneous impact at $x=0$ and at time $t_i=0.22\,T_F$ for typical superwalker parameter values. The wave fields from the \cite{Molacek2013DropsTheory} model (green dashed-dotted curve), \cite{tadrist_shim_gilet_schlagheck_2018} model (maroon dotted curve) and the superwalker model of this work (blue solid curve) are shown at times (a) $t_1=t_i+0.23\,T_F$, (b) $t_2=t_i+0.57\,T_F$, (c)  $t_3=t_i+0.76\,T_F$ and (d) $t_4=t_i+1.00\,T_F$. The evolution of the absolute wave height $h$ at $x=0$ from an impact at $t_i$ (vertical red dashed line) is shown in (e) and the relative wave height $\Delta h/h_*^{(T)}$ with respect to the \cite{tadrist_shim_gilet_schlagheck_2018} model is shown in (f). Here $h_*^{(T)}$ is the wave field from \cite{tadrist_shim_gilet_schlagheck_2018} model in \eqref{tadrist_wave} excluding the cosine term to avoid singularities in $\Delta h/h_*^{(T)}$, and $\Delta h = h^{(SW)} - h^{(T)}$ or $h^{(M)} - h^{(T)}$. The parameters are $\Gamma_{80}=3.8$, $\Gamma_{40}=0.6$ and $\Delta\phi=130^{\circ}$.}
\label{fig: wavefield}
\end{figure}

Comparing the superwalker wave field in \eqref{sw_wave} to that of a walker in \eqref{tadrist_wave} leads to two key observations: (i) both models have a wave at frequency $f/2=40$\,Hz. We note that \citet{tadrist_shim_gilet_schlagheck_2018} derived \eqref{tadrist_wave} by considering a cosine form of driving while we have considered a sine form of driving to be consistent with experiments of \citet{superwalker}. This results in a constant shift of $\pi/4$ in the phase shift $\theta^{+}_F$ in \eqref{tadrist_wave} which has been taken into account when comparing results. (ii) An additional wave of frequency $f/4=20$\,Hz appears in the wave field of a superwalker.  However, in the region of $(\Gamma_{80},\Gamma_{40})$ parameter space where superwalking is realised, typically the amplitude of the $40\,$Hz wave, $A_{40}$, is $4$ to $8$ times that of the $20\,$Hz wave, $A_{20}$.  Thus in general, our new two-frequency wave model is not appreciably different from the single-frequency model of \cite{tadrist_shim_gilet_schlagheck_2018}.  This is illustrated further in figure~\ref{fig: wavefield} where the wave fields predicted using the models of \citet{Molacek2013DropsTheory} in \eqref{molacek wave}, \citet{tadrist_shim_gilet_schlagheck_2018} in \eqref{tadrist_wave} and the superwalker wave field in \eqref{sw_wave} are shown for an instantaneous impact at time $0.22\,T_F$, corresponding to the typical impact phase for superwalkers, with an appreciable $\Gamma_{f/2}$ component. The waves from our new two-frequency model \eqref{sw_wave} and the single-frequency \cite{tadrist_shim_gilet_schlagheck_2018} model \eqref{tadrist_wave} are quantitatively similar (figures~\ref{fig: wavefield}(a) and (c-e)), except near times when the overall wave field is quite flat and is changing rapidly (figure~\ref{fig: wavefield}(b)). The comparison with the single-frequency \cite{Molacek2013DropsTheory} model appears poorer, however the difference is primarily in the far-field and arises from the absence of diffusive spatial decay in this model. In the near-field region of primary interest for walking, all three models are quantitatively similar with a maximum relative error of around $20\%$ as shown in figures~\ref{fig: wavefield}(e) and (f). Moreover, as shown in figure~\ref{fig: wavefield}(f), the relative height difference at the impact location between the waves of \citet{Molacek2013DropsTheory} model and \citet{tadrist_shim_gilet_schlagheck_2018} model is sinusoidal due to the added phase shift of $\theta_F^{+}\approx -4^{\circ}$ in the \citet{tadrist_shim_gilet_schlagheck_2018} model for the chosen parameters in figure~\ref{fig: wavefield}, and the relative height difference at the impact location between the superwalker wave and that of \citet{tadrist_shim_gilet_schlagheck_2018} reveals the added $20$\,Hz wave in the superwalker wave field. Overall these results suggest that the wave fields observed for two-frequency and single-frequency driving remain very similar, an observation that was also made qualitatively by \cite{superwalker}.  In \S\:\ref{ES}, we present results using our new two-frequency model, but note that results using either the \cite{Molacek2013DropsTheory} model or the \cite{tadrist_shim_gilet_schlagheck_2018} model are comparable; we provide details in Appendix \ref{compare models}.

\subsection{Numerical method and parameter values}\label{nmpv}

We solved \eqref{eq: vertical} and \eqref{eq: hor} using the Leap-Frog method \citep{Sprott}, a modified version of the Euler method where the new horizontal and vertical positions of the droplet are calculated using the old velocities and then the new velocities are calculated using the new positions. To increase computational speed, we only stored the waves generated by the $100$ most recent bounces of the droplet and discard the earlier ones, which have typically decayed to below $10^{-5}$ of their initial amplitude. The simulations were initialised with $\mathbf{x}_d=(0,0)\,\text{mm}$, $\dot{\mathbf{x}}_d=(1,0)\,\text{mm/s}$, $\dot{z}_d=0$\,mm/s and six different equally spaced vertical positions $z_d= (0,2,4,6,8,10)R$. Multiple initial conditions were used so that different modes existing at the same parameter values are likely to be captured. 

The physical parameters were fixed to match the experiments of \citet{superwalker}: $\rho=950$\,$\text{kg}/\text{m}^3$, $\nu=20$\,cSt, $\sigma=20.6$\,$\text{mN}/\text{m}$ and $f=80$\,Hz. There are three adjustable parameters in the model: the spring constant of the bath $k_s$, the damping coefficient of the bath $b$ and the dimensionless contact drag coefficient $C$. The dimensionless parameters corresponding to $k_s$ and $b$ are $K=k_s/m\omega_d^2$ and $B=b/m\omega_d$, where $\omega_d=\sqrt{\sigma/\rho R^3}$ is the droplet's characteristic internal vibration frequency \citep{molacek_bush_2013}. For walkers, appropriate values were determined by fitting to experimental data \citep{molacek_bush_2013,Molacek2013DropsTheory} and typical values are $K=0.59$ and $B=0.48$ \citep{couchman_turton_bush_2019}, and $C=0.17$ \citep{Molacek2013DropsTheory}. For superwalkers, we also set $C=0.17$, but adjust $K$ and $B$ to best fit the available experimental data.  The details of this fit are described in Appendix~\ref{KBR}.  We use both constant values of $K=0.70$ and $B=0.60$, as well as allowing the parameter $K$ to vary with droplet radius $R$ according to
\begin{equation}\label{kb_linear}
K=1.06\sqrt{\text{Bo}}+0.37    
\end{equation}
with a fixed $B=0.60$, where $\text{Bo}=\rho g R^2/\sigma$ is the Bond number of the droplet. We refer the reader to Appendix \ref{KBR} for more details on how this relationship was obtained. We note that these values give a good match with the experiments of \citet{superwalker}; however, the qualitative behaviour of the results remains unchanged for a range of $K$ and $B$ values.

\subsection{Description of vertical bouncing modes} \label{nomenclature}

The vertical bouncing dynamics are crucially important for the existence and characteristics of superwalkers.  Here we provide a description of the nomenclature we use to distinguish the different vertical bouncing modes.  

We follow \cite{superwalker} and use the notation $(l,m,n)$ to indicate that the droplet impacts the surface $n$ times during $m$ oscillation periods of the bath at frequency $f$, which equals $l$ oscillation periods of the bath at frequency $f/2$.  For single-frequency driving, the index $l$ is dropped.  For normal walking droplets, one of the most commonly observed mode is $(2,1)$, with the droplets leaping over every second peak in the bath's motion.  After \cite{molacek_bush_2013}, we distinguish two different styles of $(2,1)$ walking and corresponding $(1,2,1)$ superwalking, with a high-bouncing, short-contact mode denoted by $(2,1)^\text{H}$ and $(1,2,1)^\text{H}$, and a low-bouncing, long-contact mode denoted by $(2,1)^\text{L}$ and $(1,2,1)^\text{L}$. Droplets that have two peaks in the normal force during contact are classified as \otol{} while those that have only one peak are classified as \otoh{} \citep{galeano-rios_milewski_vanden-broeck_2019}. Another commonly observed mode is $(2,2)$ and corresponding $(1,2,2)$, in which the droplets no longer are able to leap over intermediate peaks and contact the bath twice, typically a high bounce and a low bounce, every two up-and-down cycles of the bath. Note that experimentally it is difficult to distinguish between a $(2,1)^\text{L}$ and $(2,2)$ mode (see figures~7 and 8 of \citep{galeano-rios_milewski_vanden-broeck_2019}).  A less commonly observed mode is $(4,2)$ and corresponding $(2,4,2)$, in which the droplets leap over every second peak, but each bounce in a pair has different amplitude. Finally, bouncing modes with no discernible periodicity or those with periodic contact but aperiodic modulation of peak bouncing heights are referred to as chaotic modes. 

\section{Emergence of superwalking}\label{ES}

\begin{figure}
\centering
\includegraphics[width=\columnwidth]{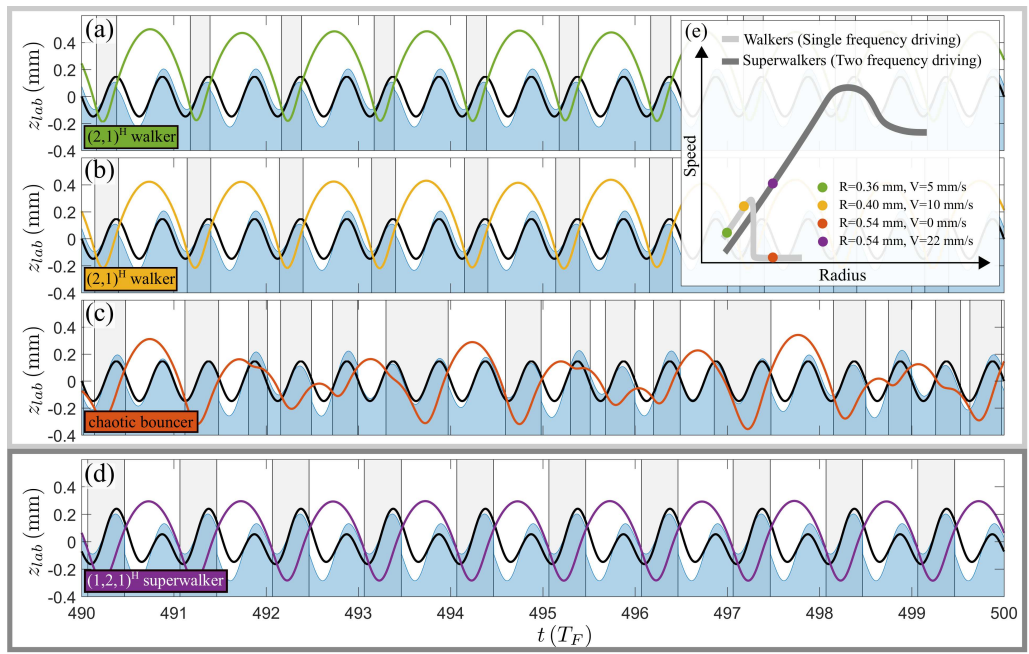}
\caption{Emergence of a superwalker. Panels (a)-(c): Vertical dynamics of a walker of radius (a) $R=0.36$\,mm and (b) $R=0.40$\,mm bouncing in a $(2,1)^\text{H}$ mode, and a bigger non-walking droplet of radius (c) $R=0.54$\,mm  bouncing in a chaotic mode. Here the bath is driven at a single frequency of $f=80$\,Hz with acceleration amplitude $\Gamma_{80}=3.8$. Panel (d): Vertical dynamics of a superwalker of radius $R=0.54$\,mm bouncing in a \otoh{} mode. Here the bath is driven at $f=80$\,Hz and $f/2=40$\,Hz with phase difference $\Delta\phi=130^{\circ}$ and acceleration amplitudes $\Gamma_{80}=3.8$ and $\Gamma_{40}=0.6$. In panels (a)-(d), the solid black curves indicate the bath motion, $\mathcal{B}(t)=-(\Gamma_f\text{g}/\Omega^2)\sin(\Omega t)-(4\Gamma_{f/2}\text{g}/\Omega^2)\sin(\Omega t/2+\Delta\phi)$, the coloured curves represent the motion of the south pole of the droplet, $z_d(t)+\mathcal{B}(t)$, and the filled blue regions illustrate the motion of the liquid surface, $h(\mathbf{x}_d,t)+\mathcal{B}(t)$, all in the lab frame. The grey regions indicate times at which the droplet is in contact with the bath. Panel (e) shows a schematic of the speed-size characteristics for the droplets in panels (a)-(d). Here the values of the parameters $K$ and $B$ are both fixed to $0.70$ and $0.60$ respectively.}
\label{fig: sw_compare}
\end{figure}

To illustrate the emergence of superwalking and its relationship with normal walking, we begin by describing the dynamics of a relatively small normal walker with the bath driven at a single frequency of $f=80$\,Hz and acceleration amplitude $\Gamma_{80}=3.8$ (compared to a Faraday threshold $\Gamma_{F80}=4.15$). This results in a normal walker that is bouncing in a $(2,1)^{\text{H}}$ mode (see figure~\ref{fig: sw_compare}(a)). The $(2,1)$ bouncing mode is crucial for walking as the droplet is bouncing at the same frequency as the frequency of the subharmonic Faraday waves that emerge beyond the Faraday instability threshold. Thus the droplet's bouncing is in resonance with the damped Faraday waves it generates and with which it interacts. For slightly larger droplets (see figure~\ref{fig: sw_compare}(b)), the same $(2,1)^{\text{H}}$ bouncing mode is maintained but the height of the bounces are reduced, while for larger droplets still, the bounces reduce in height to such an extent that the droplet can no longer leap over the second peak in the bath's motion. For the chosen parameters, this results in the droplet transitioning to a chaotic bouncing mode and no longer walking (figure~\ref{fig: sw_compare}(c)). 
 
In contrast, figure~\ref{fig: sw_compare}(d) shows the vertical dynamics of the same-sized droplet as in figure~\ref{fig: sw_compare}(c) with the addition of the subharmonic frequency $f/2=40$\,Hz and amplitude $\Gamma_{40}=0.6$ (compared to a Faraday threshold $\Gamma_{F40}=1.22$) at a phase difference of $\Delta\phi=130{^\circ}$. This additional subharmonic driving raises every second peak and lowers the intermediate peaks in both the bath's and the waves' motion. This allows the bigger droplet to clear every second peak in the bath's motion and settle in a \otoh{} bouncing mode, effectively identical to the $(2,1)^{\text{H}}$ mode for a walker, and results in the emergence of a superwalker. This jump from a walker to a superwalker is shown schematically on the speed-size curve in figure~\ref{fig: sw_compare}(e).

\begin{figure}
\centering
\includegraphics[width=\columnwidth]{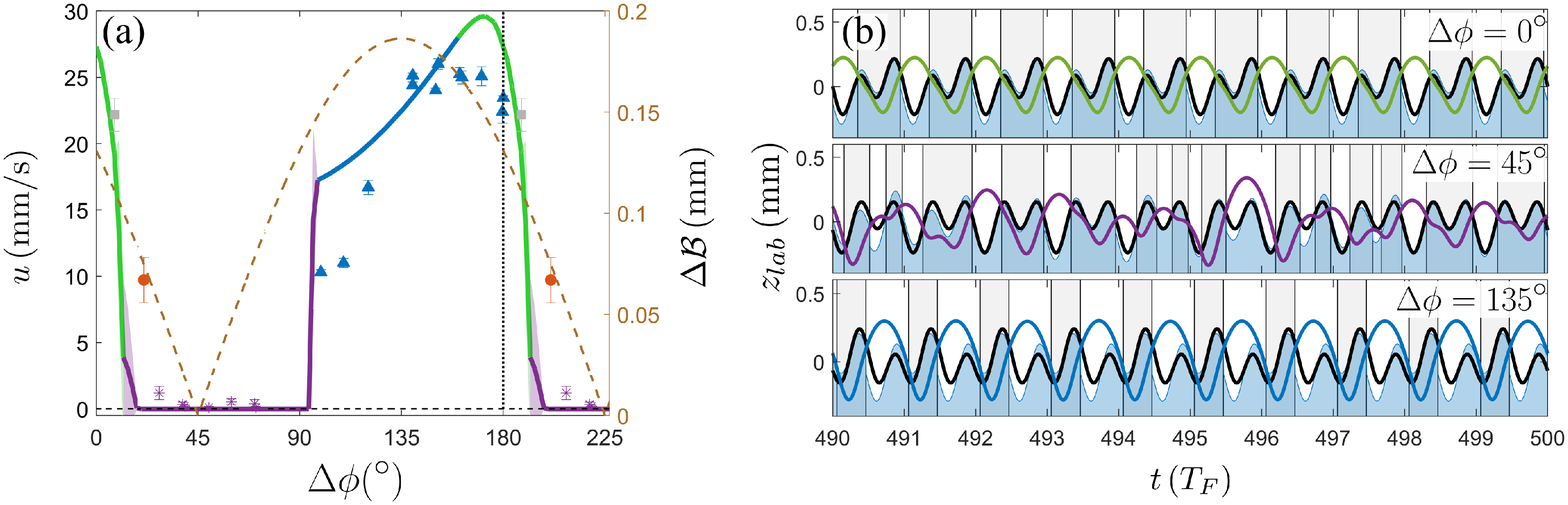}
\caption{Effect of phase difference on superwalking behaviour. (a) Walking speed $u$ as a function of the phase difference $\Delta\phi$ for a superwalker of radius $R=0.54$\,mm with $\Gamma_{80}=3.8$ and $\Gamma_{40}=0.6$. The solid curve represents results from numerical simulations with colours indicating different bouncing modes: \otol{} in green, \otoh{} in blue, and chaotic in purple. The experimental results of \citet{superwalker} are shown by points, with the style of marker indicating the bouncing modes: $(1,2,2)$ are red circles $\bullet$, \otoh{} are blue triangles {\protect \scalebox{0.8}{$\blacktriangle$}}, transition between a \otoh{} and a $(1,2,2)$ mode are grey squares {\protect \scalebox{0.5}{$\blacksquare$}}, and chaotic are purple asterisks $*$. The dashed curve indicates the height difference 
$\Delta \mathcal{B}$ between consecutive peaks in one period of the bath motion. The data to the right of the vertical dotted line is repeated. Panel (b) shows bouncing modes obtained for different values of $\Delta\phi$ from panel (a). In this panel, the grey regions indicate times at which the droplet is in contact with the bath. The parameters $K$ and $B$ are fixed to $0.70$ and $0.60$ respectively.}
\label{fig: Phasespeed}
\end{figure}

\subsection{Importance of the phase difference between the two driving frequencies}\label{Phase}

The phase difference between the two driving frequencies controls the relative height of the two peaks in one full cycle of the periodic bath motion, equivalently two up-and-down cycles, and it is therefore a crucial parameter for the emergence of superwalking. Figure~\ref{fig: Phasespeed}(a) shows the walking speed $u$ as a function of the phase difference $\Delta\phi$ for a fixed-sized droplet that is too large to walk with single-frequency driving (the largest droplet shown in figure~\ref{fig: sw_compare}). The different vertical modes at different $\Delta\phi$ are shown in figure~\ref{fig: Phasespeed}(b). Depending on the phase difference, the droplet either bounces without walking or it superwalks. In the bouncing regime, $20^{\circ}\lesssim\Delta\phi\lesssim90^{\circ}$, the droplet's vertical dynamics appear chaotic. This can be attributed to the height difference $\Delta\mathcal{B}$ between successive peaks in the bath's motion being small (see dashed curve in figure~\ref{fig: Phasespeed}(a)) and hence the droplet behaves similarly to the single frequency case (see figure~\ref{fig: sw_compare}(c)). Conversely, regions of high superwalking speed are associated with a large height difference $\Delta\mathcal{B}$ between the two peaks in the bath's motion and a droplet can easily settle in a $(1,2,1)$ bouncing mode. 

The predicted speeds from the numerical simulations agree well with experiments.  The chaotic mode in the bouncing regime and the \otoh{} bouncing mode in the superwalking regime are also observed at parameter values comparable to those in experiments.  The numerically observed \otol{} superwalkers were not reported in experiments, instead $(1,2,2)$ modes were observed at the corresponding parameter values.  However, as noted in \S\:\ref{nomenclature}, it is difficult to distinguish between a $(1,2,1)^\text{L}$ and a $(1,2,2)$ mode experimentally. Hence, it is not clear whether all the $(1,2,2)$ superwalkers reported by \cite{superwalker} are truly $(1,2,2)$ superwalkers or if some may in fact be \otol{} superwalkers.

\subsection{Speed-size characteristics of superwalking droplets}\label{SpeedSize}

\begin{figure}
\centering
\includegraphics[width=\columnwidth]{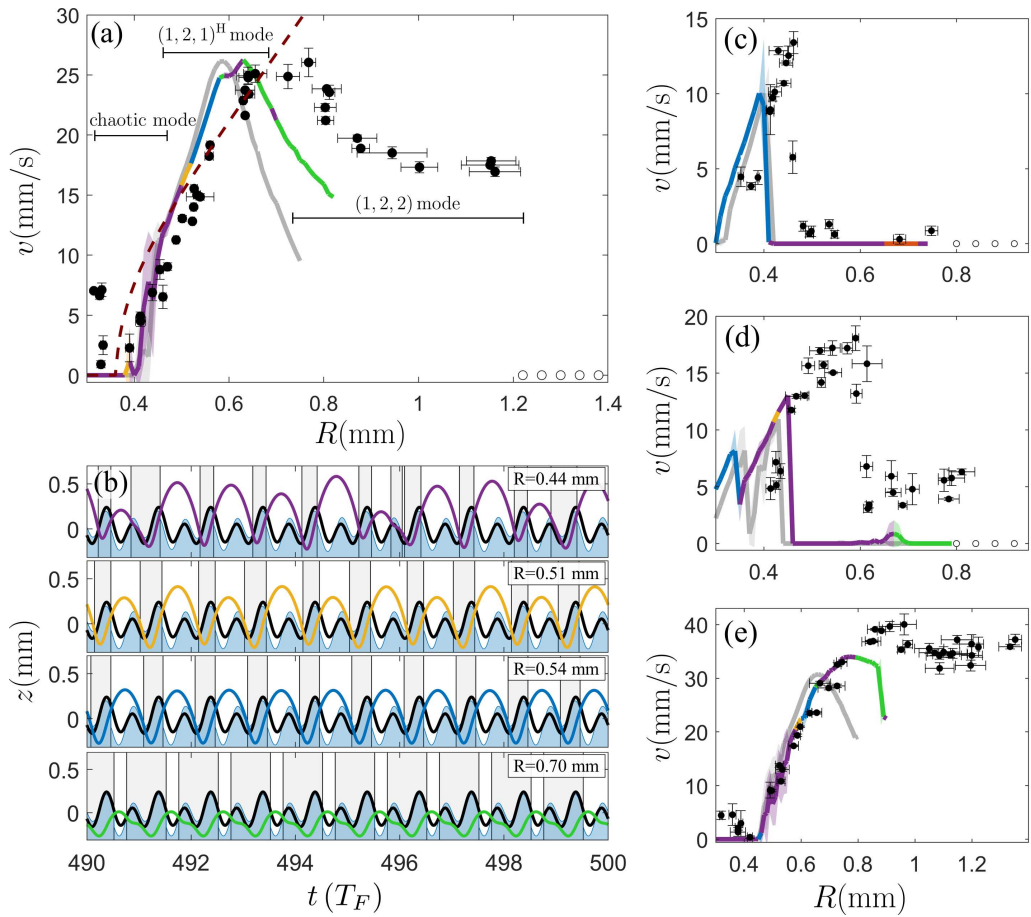}
\caption{Speed of a superwalker as a function of its size at fixed  $\Gamma_{80}=3.8$ and $\Delta\phi=130^{\circ}$. (a) Comparison of the speed-size characteristics of a droplet from numerical simulations (solid curves) with experimental results of \citet{superwalker} (black circles with empty circles indicating coalescence) and the stroboscopic model of \citet{Oza2013} (dashed curve) at $\Gamma_{40}=0.6$. For the stroboscopic model, we set the adjustable parameters of the impact phase and the non-dimensional drag coefficient to
$\sin(\Phi) = 0.2$ and $C = 0.17$ respectively, while the other parameters were chosen to match the experiments of \citet{superwalker}.  The black horizontal bars indicate where different bouncing modes in experiments were observed. Panel (c), (d) and (e) show the speed-size characteristics at $\Gamma_{40}=0$, $\Gamma_{40}=0.3$ and $\Gamma_{40}=1$ respectively. In each panel the grey curve is for fixed parameter values of $K=0.70$ and $B=0.60$, and multicoloured curve represents when $K$ is varied as a linear function of the droplet radius $R$ according to \eqref{kb_linear} with a fixed $B=0.60$. The colours on this curve represent a chaotic bouncing mode in purple, $(2,4,2)$ mode in yellow, \otoh{} mode in blue, \otol{} mode in green and $(1,2,2)$ mode in red. Termination of the solid curves indicate coalescence. The typical bouncing modes from panel (a) at different droplet radii are shown in panel (b). In this panel, the grey regions indicate times at which the droplet is in contact with the bath.}  
\label{fig: SpeedSize}
\end{figure}

In the size range for which walkers exist, their walking speed typically increases with their size \citep{Molacek2013DropsTheory}. For superwalkers, two trends were observed in experiments: an ascending branch for smaller superwalkers where the speed increases with size, followed by a descending branch for larger superwalkers where the speed decreases with size \citep{superwalker}. Figure~\ref{fig: SpeedSize} shows the speed-size characteristics of simulated superwalkers at $\Gamma_{80}=3.8$ and $\Delta\phi=130^{\circ}$ for a range of $\Gamma_{40}$ values. 

We begin by focusing on the comparison for the ascending branch. Simulated superwalking speeds for different droplet radii for constant $K=0.70$ and $B=0.60$ (grey curves) as used in the simulations shown in figures~\ref{fig: sw_compare} and~\ref{fig: Phasespeed}, and $K$ linearly increasing function of droplet radius as in \eqref{kb_linear} with a fixed $B=0.60$ (coloured curves) are shown. We refer the reader to Appendix \ref{KBR} for details on this linear relationship. Both the superwalking speed and the bouncing mode are captured well for both combinations for small- to moderate-sized superwalkers, and this is generally true for a broad range of $K$ and $B$ values (see Appendix~\ref{KBR}). By allowing $K$ to vary linearly with the droplet radius $R$ (coloured curve), we obtain a better fit for droplets on the ascending branch at relatively high $\Gamma_{40}$ values (see figure~\ref{fig: SpeedSize}(e)). Focusing on the vertical dynamics for this fit when $\Gamma_{40}=0.6$ (see figure~\ref{fig: SpeedSize}(a)), we find that superwalkers on this branch universally impact the bath once every two up and down cycles of the bath's motion.  For the smallest superwalkers, the amplitude of the bounces is chaotic. As the droplet size increases, there is a transition to a $(2,4,2)$ mode in a narrow region near $R=0.51$\,mm.  Beyond this, the droplets bounce in a \otoh{} mode (blue) for the remainder of the ascending branch. This agrees well with the experimental results of \citet{superwalker} where chaotic and \otoh{} bouncing modes were also reported on the ascending branch. 

Simulations of larger droplets that lie on the descending branch in experiments reveal that the model is unable to capture the larger superwalkers. We have explored different constant values of $K$ and $B$ as well as varying $K$ and $B$ as a function of $R$ but were unable to obtain a better fit to the experimental superwalking speeds on this branch than the relatively poor fits shown in figure~\ref{fig: SpeedSize}. However, we note that the bouncing modes predicted from simulations on the descending branch are comparable with experimental observations. For the curves shown, the superwalkers on the descending branch bounce typically bounces in a \otol{} mode. Although only the $(1,2,2)$ mode was reported in experiments, as previously mentioned, \otol{} and $(1,2,2)$ are similar and have been difficult to distinguish in experiments. 

\subsection{Superwalking behaviour in the $(\Gamma_{80},\Gamma_{40})$ parameter space}

\begin{figure}
\centering
\includegraphics[width=\columnwidth]{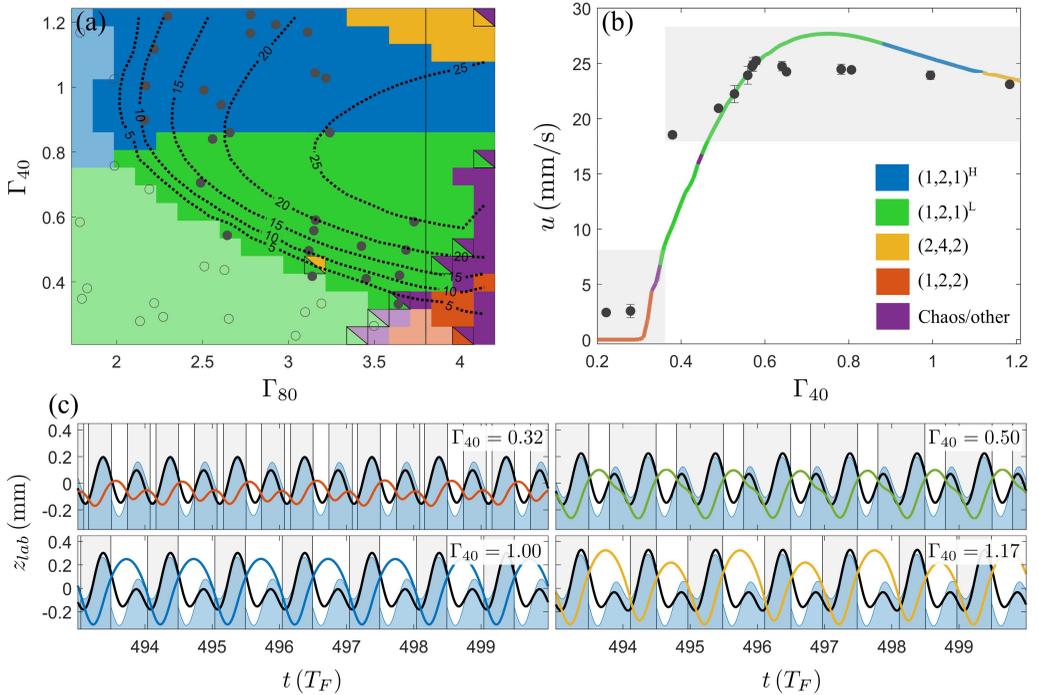}
\caption{Superwalking behaviour in the $(\Gamma_{80}$,$\Gamma_{40})$ parameter space. (a) Bouncing modes shown as different colours for a droplet of radius $R=0.60$\,mm in the $(\Gamma_{80}$,$\Gamma_{40})$ parameter space with multiple colours at the same $(\Gamma_{80}$,$\Gamma_{40})$ values indicating multiple bouncing modes that were observed at the same $(\Gamma_{80}$,$\Gamma_{40})$ values. The lighter shade of each colour indicates bouncing and the darker shade is where superwalking is observed with walking speed indicated by dotted constant speed contours in mm/s. The markers indicate bouncing (empty circles) and superwalking (filled circles) for a droplet of radius $R=(0.60\pm0.05)$\,mm from the experiments of \citet{superwalker}. (b) A vertical slice of the parameter space in panel (a) (solid line) showing walking speed $u$ as a function of $\Gamma_{40}$ at a fixed $\Gamma_{80}=3.8$. The solid curve is the result from simulations with colours indicating bouncing modes and the filled black markers are the experimental walking speeds from \citet{superwalker} for a droplet of radius $R=(0.63\pm0.03)$\,mm. The grey shaded region indicates the jump in walking speed for this droplet when $\Gamma_{40}$ is appreciable. The different bouncing modes at different $\Gamma_{40}$ values are shown in panel (c) with the grey regions in this panel indicating contact with the bath. The phase difference is fixed to $\Delta\phi=130^{\circ}$. The parameters $K$ and $B$ are fixed to $0.70$ and $0.60$ respectively.}
\label{fig: PS}
\end{figure}

By fixing the phase difference $\Delta\phi$ and the droplet radius $R$, we can explore the vertical and horizontal dynamics of a droplet in the parameter space formed by the two acceleration amplitudes $\Gamma_{80}$ and $\Gamma_{40}$. We choose a droplet radius of $R=0.60$\,mm and a phase difference of $\Delta\phi=130^{\circ}$ to compare our results with the experimental results of \citet{superwalker}. Figure~\ref{fig: PS}(a) shows the region of parameter space where bouncing (lighter shades) and superwalking (darker shades) are observed as well as the bouncing modes (different colours) observed in those regions. Regions of bouncing (empty circles) and superwalking (filled circles) that were identified in the experiments of \citet{superwalker} are also shown. We find an excellent agreement in the transition boundary from bouncing to superwalking. Moreover, we identify that the superwalking region is dominated by the $(1,2,1)$ bouncing mode with a \otol{} mode when $\Gamma_{40}$ is small and a \otoh{} mode when $\Gamma_{40}$ is large. In contrast, the bouncing mode is nearly independent of $\Gamma_{80}$ at a fixed $\Gamma_{40}$ except at relatively high $\Gamma_{80}$ values.

To understand how the superwalking speed $u$ changes as a function of $\Gamma_{40}$, we show a slice of the $(\Gamma_{80},\Gamma_{40})$ parameter space at $\Gamma_{80}=3.8$ in figure~\ref{fig: PS}(b) with corresponding bouncing modes in figure~\ref{fig: PS}(c). We find that the walking speed is initially zero for all $\Gamma_{40}\lesssim0.3$ before increasing rapidly with $\Gamma_{40}$ to a peak value near $\Gamma_{40}=0.7$ and then marginally decreasing.  This illustrates the rather abrupt rise in walking speed that occurs once the asymmetry between the heights of succeeding peaks in the bath's and waves' motion is sufficient. Comparison of this numerically simulated walking speed $u$ versus acceleration amplitude $\Gamma_{40}$ curve with that obtained from the experiments of \citet{superwalker} for a droplet radius of $R=(0.63\pm0.03)$\,mm, shows good agreement highlighting the success of the present model for small- to moderate-sized superwalkers. 

\section{Discussion and conclusions}\label{section: DC}

We have studied the dynamics of bouncing droplets on a vibrating liquid bath under two-frequency driving using the theoretical model of \citet{molacek_bush_2013,Molacek2013DropsTheory} and a new model for the wave field to understand the emergence of superwalkers. We have shown that two-frequency driving at $f$ and $f/2$ with an appropriately chosen phase difference $\Delta\phi$ lifts every second peak and lowers the intermediate peaks in the bath's motion. This allows larger droplets to bounce in a resonant $(1,2,1)$ mode where they can efficiently excite damped subharmonic Faraday waves that enable them to superwalk. We note that superwalking would not be expected for two arbitrary frequency combinations, as the lowering of every second peak is crucial for droplets to remain in a $(1,2,1)$ mode. For example, for two frequency driving at $f=80$\,Hz and $4f/5=64$\,Hz, \citet{PhysRevE.94.053112} reported chaotic bouncing modes with irregular walking at typical speeds of only $5\,\text{mm/s}$.

We have shown that the phase difference $\Delta\phi$ plays a crucial role in the dynamics of superwalking droplets because it controls the relative amplitudes of two succeeding peaks in one full cycle of the bath's motion. Fast superwalking occurs for phase differences between $130^\circ$ and $180^\circ$ where there is a larger difference between these amplitudes, while phase differences around $45^\circ$, where the amplitude difference between succeeding peaks is small, correspond to stationary bouncing or coalescence.  Extending these observations, it will be interesting to use our model to explore a system with slightly detuned driving frequencies of $f$ and $f/2+\epsilon$ with $\epsilon\ll f/2$.  This is equivalent to driving at frequencies of $f$ and $f/2$ with a slowly varying phase difference $\Delta\phi(t)=\Delta\phi + 2\pi\epsilon t$. Such traversal of the phase difference gives rise to new types of locomotion in the walking droplet system where, \emph{e.g.}, the droplet alternates periodically between superwalking and stationary bouncing.  Such behaviour was observed experimentally by \citet{superwalker} and was coined stop-and-go motion.  We aim to discuss the details of such motion in a future work.

On comparing the speed-size characteristics of simulated superwalkers with the experimental results of \citet{superwalker}, we find excellent agreement on the ascending branch, with \otoh{} superwalkers primarily observed. These observations also explain the good agreement noted by \cite{superwalker} between superwalking speeds obtained in experiments and those predicted using the stroboscopic model of \cite{Oza2013} (dashed curve in figure~\ref{fig: SpeedSize}(a)). The latter is a reduced form of the full \cite{molacek_bush_2013,Molacek2013DropsTheory} model predicated on a $(2,1)^\text{H}$ bouncing mode and our two-frequency model would reduce to essentially the same model for such modes. 

The superwalking speed of larger superwalkers is not captured well by the current model. This suggests that the model does not include the fundamental mechanism that allows the largest superwalkers to walk, and even exist. Indeed, \citet{superwalker} noted that the largest superwalkers on the descending branch undergo significant internal deformations \citep{superwalker}. In Appendix~\ref{compare models}, we present results incorporating deformation of the droplets by modelling them as a vertical spring following \citet{deformationdroplet} and \citet{Gilet_2008}, and find this to have a limited effect on the speed-size curve. We also present results using the nonlinear logarithmic spring model of \citet{Molacek2013DropsTheory} for the vertical dynamics with no better success. Another observation made by \cite{superwalker} was that larger superwalkers have a prolonged contact time with the bath. This prolonged contact time, potentially in combination with internal deformation, may change the wave field in the vicinity of the droplet and the long time approximation of the standing wave field in equation \eqref{sw_wave} may break down. Perhaps a more refined modelling of the system that incorporates the detailed contact interaction between the droplet and the bath, the wave evolution and droplet deformations might be required to capture the behaviour of these larger superwalkers. 
 
\section*{Acknowledgements}
\begin{acknowledgments}
We acknowledge financial support from an Australian Government Research
Training Program (RTP) Scholarship (R.V.) and the Australian Research Council via the Future Fellowship Project No.\ FT180100020 (T.S.). 
\end{acknowledgments}

\section*{Declaration of Interests} 
The authors report no conflict of interest.

\appendix
\section{Derivation of waves generated by a superwalker}\label{der}
To derive the form of the surface waves generated by a superwalker, we closely follow the approach of \citet{tadrist_shim_gilet_schlagheck_2018} who considered walkers driven at a single frequency. We consider an incompressible, Newtonian liquid in a bath that is infinitely large in horizontal extent and infinitely deep. The bath is subjected to periodic vertical vibrations that result in a modulation of the effective gravity in the frame of the bath $\text{g}^*(t)=\text{g}[1+\Gamma_{f} \sin(\Omega t) + \Gamma_{f/2}\sin(\Omega t/2 + \Delta \phi)]$, where $\text{g}$ is the gravitational acceleration constant and $\Gamma_f$ and $\Gamma_{f/2}$ are dimensionless acceleration amplitudes of the two frequencies with a relative phase difference of $\Delta\phi$.  For the sake of notational clarity, we will refer to specific frequencies $f=\Omega/2\pi=80$\,Hz and $f/2=40$\,Hz, but the derivation is general. The horizontal coordinates are $(x,y)$ and the vertical coordinate is $z$ with the origin located on the free surface of the undeformed liquid. The evolution of the liquid is governed by the incompressible Navier-Stokes equations,
\begin{equation*}
    \nabla \cdot \mathbf{v} = 0 \quad \text{and} \quad \frac{\partial \mathbf{v}}{\partial t} + (\mathbf{v}\cdot\nabla)\mathbf{v}=-\frac{1}{\rho}\nabla P - \text{g}^*(t)\mathbf{\hat{k}}+\nu\nabla^2\mathbf{v}, 
\end{equation*}
where $\mathbf{v}(\mathbf{r},t)$ is the velocity field, $P(\mathbf{r},t)$ is the pressure field relative to atmospheric pressure, $\mathbf{r}=(x,y,z)$ is the position vector, $\mathbf{\hat{k}}$ is a unit vector in the z direction, $\rho$ is the density and $\nu$ is the kinematic viscosity. At the free surface of the liquid, $z=h(x,y,t)$, the kinematic boundary condition implies
\begin{equation*}
    \frac{\partial h}{\partial t}+\mathbf{v}\cdot\nabla (h-z)=0,
\end{equation*}
while balancing stresses requires 
\begin{equation*}
    -P\mathbf{\hat{n}}+\rho\nu\left(\nabla\mathbf{v} +\nabla \mathbf{v}^T\right)\cdot \mathbf{\hat{n}}=-P^{\text{ext}}\mathbf{\hat{n}}-\sigma (\nabla \cdot \mathbf{\hat{n}})\mathbf{\hat{n}},
\end{equation*}
where $\sigma$ is the coefficient of surface tension, $P^{\text{ext}}(x,y,t)$ is the pressure exerted by the droplet on the free surface through the intervening air layer and $\mathbf{\hat{n}}$ is the unit normal out of the liquid. Assuming that the pressure distribution imparted by the droplet during contact is uniform in the contact region we get
\begin{equation}\label{eq: press_for}
P^{\text{ext}}(x,y,t)=P^{\text{ext}}(t)=F_N(t)/\pi w^2,
\end{equation}
where $w$ is the effective radius of the contact area and $F_N(t)$ is the normal force as described in \S\:\ref{vd}. 

We consider small perturbations from the stationary equilibrium state $\mathbf{{v}}=\boldsymbol{0}$, ${{P}}=-\rho \text{g}^*(t) z$ and $h=0$. We linearise the above equations about this equilibrium state with pressure perturbation $p(\mathbf{r},t)$,  velocity perturbation $\mathbf{v}(\mathbf{r},t)$ and free surface perturbation $h(\mathbf{r},t)$ and follow the steps outlined in \S\:2.1 of \citet{tadrist_shim_gilet_schlagheck_2018}. For the remainder of this Appendix, we will use the dimensionless time $\tau=\Omega t/2$ for ease of comparison with their equations and will revert back to the dimensional time $t$ in the final expressions.  Taking a Fourier transform in $x$ and $y$ of the linearised equations, followed by a Laplace transform and some algebra, we obtain the equation
\begin{equation}\label{twofreqF_full}
    f_k(s)h_{\mathbf{k},s}+\frac{2\Gamma_{80} \text{g} k}{\text{i}\Omega^{2}}(h_{\mathbf{k},s-2\text{i}}-h_{\mathbf{k},s+2\text{i}})+\frac{2\Gamma_{40} \text{g} k}{\text{i}\Omega^{2}}(h_{\mathbf{k},s-\text{i}}\text{e}^{i\Delta\phi}-h_{\mathbf{k},s+\text{i}}\text{e}^{-i\Delta\phi})+\Pi_{\mathbf{k},s}=0
\end{equation}
for the transformed free surface $h_{\mathbf{k},s}$.  Here the Fourier transform for an arbitrary variable $X(x,y,\tau)$ is defined as
\begin{equation*}
    X_{\mathbf{k}}(\tau) = \int_0^\infty \int_0^\infty X(x,y,\tau) \,\text{exp}[-\text{i}(k_x x + k_y y)]\,\text{d}x\,\text{d}y
\end{equation*}
with $k=|\mathbf{k}|$ and the Laplace transform is defined as
\begin{equation*}
    X_{\mathbf{k},s} = \int_0^\infty X_{\mathbf{k}}(\tau) \,\text{e}^{-s\tau}\,\text{d}\tau.
\end{equation*}
Furthermore,
\begin{equation*}
    f_k(s) = (s+\gamma_k)^2-\gamma_k^{3/2}\sqrt{\gamma_k+2s}+\omega^2_{k},
\end{equation*}
with $\gamma_k=4\nu k^2/\Omega$ and $\omega^2_k=4[\text{g} k+(\sigma/\rho)k^3]/\Omega^2$. This function and all its derivatives obey $f_k(\overline{z})=\overline{f_k(z)}$ where the overline denotes complex conjugation. 
The last term of (\ref{twofreqF_full}) described the transformed pressure distribution from droplet's impact $\Pi_{\mathbf{k},s}=(4k/\Omega^2\rho)P^{\text{ext}}_{\mathbf{k},s}$. Using the definition of Fourier transform and equation \eqref{eq: press_for} we get assuming $wk \ll 1$, $P^{\text{ext}}_{\mathbf{k},s}=P^{\text{ext}}(s)\int_0^w\text{J}_0(kr)r\text{d}r\approx F_N(s)/2\pi$ with $r=\sqrt{x^2+y^2}$. Hence we obtain
\begin{equation}\label{eq: pi_force}
\Pi_{\mathbf{k},s}=\frac{2k}{\pi\Omega^2\rho}F_N(s),    
\end{equation}
where $P^{\text{ext}}(s)$ and $F_N(s)$ are the Laplace transforms of $P^{\text{ext}}(\tau)$ and $F_N(\tau)$ respectively. We note that \eqref{twofreqF_full} reduces to equation (2.2) of \citet{tadrist_shim_gilet_schlagheck_2018} on setting $\Gamma_{40}=0$, with the caveat that our driving is a sine function while \citet{tadrist_shim_gilet_schlagheck_2018} use a cosine. We have chosen a sine function for driving for the sake of consistency with the experiments of \cite{superwalker}.

We first consider Faraday waves in the absence of external pressure perturbations, which reduces \eqref{twofreqF_full} to
\begin{equation}\label{twofreqF}
    f_k(s)h_{\mathbf{k},s}+\frac{2\Gamma_{80} \text{g} k}{\text{i}\Omega^{2}}(h_{\mathbf{k},s-2\text{i}}-h_{\mathbf{k},s+2\text{i}})+\frac{2\Gamma_{40} \text{g} k}{\text{i}\Omega^{2}}(h_{\mathbf{k},s-\text{i}}\text{e}^{i\Delta\phi}-h_{\mathbf{k},s+\text{i}}\text{e}^{-i\Delta\phi})=0.
\end{equation}

Due to the periodic driving of the system, a Floquet ansatz is appropriate in the time domain. The form we assume and its corresponding Laplace transform are given by  \citep{PhysRevE.54.507}
\begin{equation}\label{eq: flo_ana}
h_{\mathbf{k}}(\tau)=\sum_{l=-\infty}^{\infty} h^{(l)}_{\mathbf{k}}\text{e}^{\text{i}l\tau}\text{e}^{\delta_{k}\tau}\qquad\text{and}\qquad h_{\mathbf{k},s}=\sum_{l=-\infty}^{\infty} \frac{h^{(l)}_{\mathbf{k}}}{s-\text{i}l-\delta_k}.
\end{equation}
Here $\delta_k$ is a complex perturbation whose real part vanishes when the Faraday instability threshold is reached. 
Substituting this form into \eqref{twofreqF}, we obtain
\begin{align*}
\sum_{l=-\infty}^{\infty} h^{(l)}_{\mathbf{k}} \bigg[ \frac{f_k(s)}{s-\text{i}l-\delta_k} &- \text{i} \Gamma_{80}\beta_k \left( \frac{1}{s-\text{i}(l+2)-\delta_k} - \frac{1}{s-\text{i}(l-2)-\delta_k}\right) \nonumber \\
& - \text{i} \Gamma_{40} \beta_k \left( \frac{1}{s-\text{i}(l+1)-\delta_k}\text{e}^{\text{i}\Delta\phi} - \frac{1}{s-\text{i}(l-1)-\delta_k}\text{e}^{-\text{i}\Delta\phi}\right) \bigg]=0,
\end{align*}
with $\beta_k=2\text{g}k/\Omega^2$. Using the Heaviside cover-up method \cite[]{thomas_finney_1996} yields an infinite dimensional linear system $\mathsfbi{A} \mathbf{h}=\boldsymbol{0}$ coupling the Floquet components together \citep{kumar_tuckerman_1994,kumar1996,PhysRevE.54.507,tadrist_shim_gilet_schlagheck_2018}. Here $\mathsfbi{A}$ is the pentadiagonal matrix 
\begin{gather*}\label{inf_me_sin}
 \mathsfbi{A} = 
 \begin{bmatrix}
 \ddots & \ddots & \ddots & \vdots & \vdots & \vdots & \dots \\
 \ddots & f_k(-2\text{i}+\delta_k) & \ups & \alp & 0 & 0 & \dots \\ 
 \ddots & \bups & f_k(-1\text{i}+\delta_k) & \ups & \alp & 0 & \dots\\
 \dots & \balp & \bups & f_k(\delta_k) & \ups & \alp & \dots\\
 \dots & 0 & \balp & \bups & f_k(1\text{i}+\delta_k) & \ups & \ddots\\
 \dots & 0 & 0 & \balp & \bups & f_k(2\text{i}+\delta_k) & \ddots\\
 \dots & \vdots & \vdots & \vdots & \ddots & \ddots & \ddots 
 \end{bmatrix},
\end{gather*}
with $\alp=\text{i}\Gamma_{80} \beta_k$ and $\ups=\text{i}\Gamma_{40}\beta_k\text{e}^{-\text{i}\Delta\phi}$, and $\mathbf{h}$ is a vector of the Floquet components $h^{(l)}_{\mathbf{k}}$. To obtain non-trivial solutions of this linear system requires
\begin{equation}\label{inf_det}
    \text{det}(\mathsfbi{A})=0.
\end{equation}

\subsection{Decay rate of damped Faraday waves and the Faraday instability threshold}

\begin{figure}
\centering
\includegraphics[width=\columnwidth]{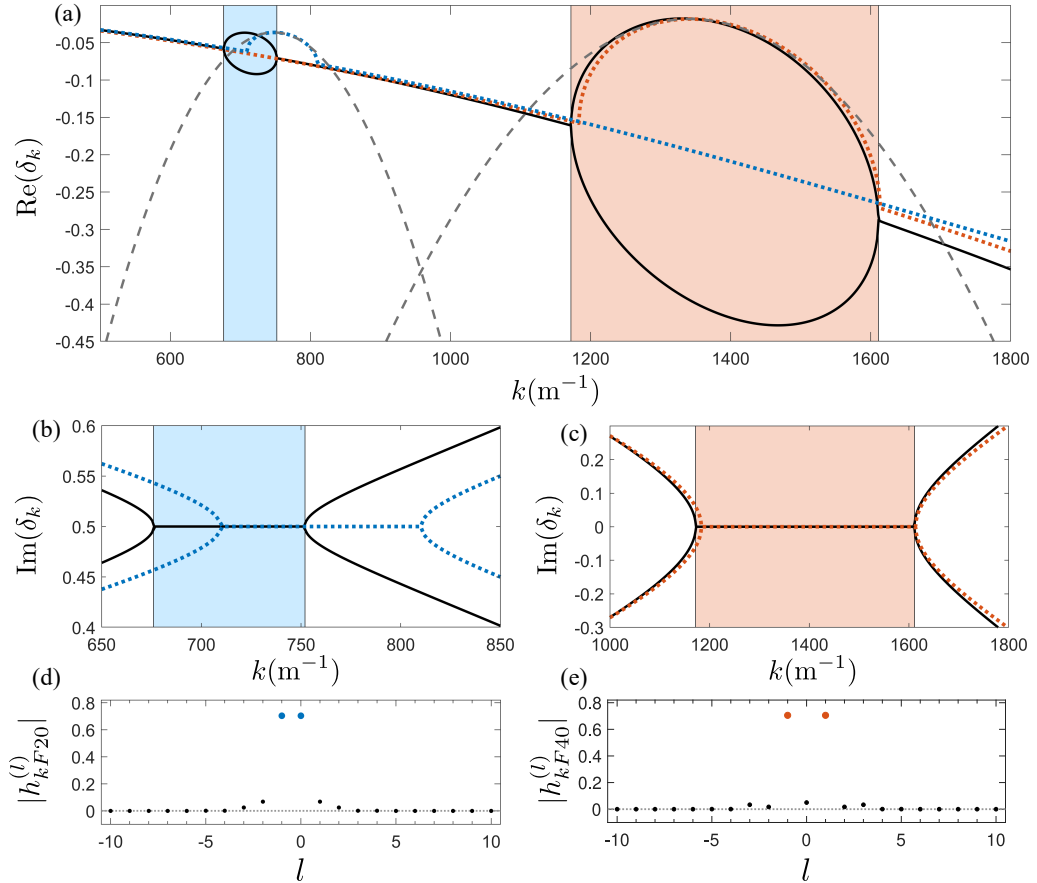}
\caption{Properties of two-frequency, damped Faraday waves. (a) Decay rate $\text{Re}(\delta_k)$ as a function of wavenumber $k$ for $\Gamma_{80}=3.8$, $\Gamma_{40}=0.6$ and $\Delta\phi=130^{\circ}$ using a $21$-mode truncation corresponding to $|l|\le 10$ (solid black curve). The blue and red dotted curves show the decay rate of the slowly decaying wave using the two-mode approximation $\text{Re}(\delta^{+}_{k20})$ in the blue Faraday window and the two-mode approximation $\text{Re}(\delta^{+}_{k40})$ in the red Faraday window, respectively. The grey dashed curves are second-order approximations to these decay rates at the peak values in each Faraday window. Panels (b) and (c) show the dispersion relation $\text{Im}(\delta_k)$ in the two Faraday windows. In (d) and (e), the magnitude of the amplitudes $h_{kF40}^{(l)}$ and $h_{kF20}^{(l)}$ of the different modes $l$ are shown at the most unstable wavenumbers in each Faraday window using the $21$-mode truncation, with the dominant modes coloured. These correspond to the eigenvectors of $\mathsfbi{A} \mathbf{h}=\boldsymbol{0}$ with eigenvalue $0$. Note that these amplitudes only yield information about the relative values of each mode.}
\label{fig: dk_plots}
\end{figure}

Solving \eqref{inf_det} for fixed $\Gamma_{80}$, $\Gamma_{40}$ and $\Delta\phi$, we obtain $\delta_k$ as a function of the wavenumber $k$. Below the Faraday instability threshold, this corresponds to a decay rate for the waves $\text{Re}(\delta_k)$ and a dispersion relation $\text{Im}(\delta_k)$. Results for typical parameter values of superwalkers are shown in figure~\ref{fig: dk_plots}. Figure~\ref{fig: dk_plots}(a) shows the numerically converged $\text{Re}(\delta_k)$ as a function of $k$ (solid curves). We see two different Faraday windows, one in which the waves are locked at $\text{Im}(\delta_k)=1/2$ (the blue-shaded region in figures~\ref{fig: dk_plots}(a) and (b)) and one in which waves are locked at $\text{Im}(\delta_k)=0$ (the red-shaded region in figures~\ref{fig: dk_plots}(a) and (c)). In each of these  windows, we see two different branches for the decay rate $\text{Re}(\delta_k)$, an upper branch corresponding to a slowly decaying wave and a lower branch corresponding to a more rapidly decaying wave.

To obtain analytical forms of these results, we truncate the infinite dimensional matrix equation to a few dominant modes. For the $\text{Im}(\delta_k)=0$ (red) Faraday window in figure~\ref{fig: dk_plots}(a), we find that the dominant contribution to the amplitude is from the two modes with $l=\pm 1$ (see figure~\ref{fig: dk_plots}(e)) corresponding to a frequency of $\pm 40$\,Hz. Denoting the decay rate in this Faraday window by $\text{Re}(\delta_{k40})$ and using this two-mode approximation, \eqref{inf_det} reduces to
\begin{equation*}
    f_k(-\text{i}+\text{Re}(\delta_{k40}))f_k(\text{i}+\text{Re}(\delta_{k40}))-|\alp|^2=0.
\end{equation*}

 We can obtain a good approximation to this decay rate by following an approach similar to \S\:2.2.2 of \citet{tadrist_shim_gilet_schlagheck_2018} and expanding the function $f_k(\pm\text{i}+\text{Re}(\delta_{k40}))$ to second order in the small decay rate $\text{Re}(\delta_{k40})$ to get
\begin{equation}\label{dk_40_app}
    \text{Re}(\delta_{k40}^{\pm})=-\frac{b_{k}(\text{i})}{2a_{k}(\text{i})}\left(1 \mp \sqrt{1-\frac{4a_{k}(\text{i})c_{k}(\text{i},\alp)}{b^2_{k}(\text{i})}} \right),
\end{equation}
where the functions
\begin{align*}
a_k(u)&=\dot{f}_k(u)\dot{f}_k(-u)+\tfrac{1}{2}\ddot{f}_k(u)f_k(-u)+\tfrac{1}{2}\ddot{f}_k(-u)f_k(u),\\
b_k(u)&=\dot{f}_k(u)f_k(-u)+\dot{f}_k(-u)f_k(u),\\
c_k(u,Z)&=f_k(u)f_k(-u)-|Z|^2.
\end{align*}

Here $\text{Re}(\delta_{k40}^{+})$  and $\text{Re}(\delta_{k40}^{-})$ correspond to the decay rates of the slowly and quickly decaying wave respectively. This approximation for the slowly decaying wave $\text{Re}(\delta_{k40}^{+})$ is shown as a red, dotted curve in figure~\ref{fig: dk_plots}(a). We can further approximate this decay rate near the most unstable wavenumber $k_{F40}$ by
\begin{equation}\label{df40_eq}
    \text{Re}(\delta_{k40}^{+}) \approx \text{Re}(\delta_{F40}^{+}) - D_{40} (k-k_{F40})^2,
\end{equation}
where $\text{Re}(\delta_{F40}^{+})=\lim_{k\to k_{F40}}\text{Re}(\delta_{k40}^{+})$ and $D_{40}=-\frac{1}{2}\text{d}^2\text{Re}(\delta_{k40}^{+})/\text{d}k^2|_{k=k_{F40}}$ is the diffusion coefficient, both of which can be calculated from \eqref{dk_40_app}. This approximation of $\text{Re}(\delta_{k40}^{+})$ is shown as a grey, dashed curve in figure~\ref{fig: dk_plots}(a).

We follow a similar approach to obtain an analytical expression for the decay rate in the $\text{Im}(\delta_k)=1/2$ (blue) Faraday window in figure~\ref{fig: dk_plots}(a). In this window, the dominant contribution is from the $l=-1$ and $0$ modes (see figure~\ref{fig: dk_plots}(d)), corresponding to frequencies $\pm 20$\,Hz. Using this two-mode approximation and denoting the decay rate by $\text{Re}(\delta_{k20})$, \eqref{inf_det} reduces to
\begin{align*}
f_k(-\text{i}/2+\text{Re}(\delta_{k20}))f_k(\text{i}/2+\text{Re}(\delta_{k20}))-|\ups|^2=0.
\end{align*}
A good approximation for this decay rate is obtained by expanding the function $f_k(\pm\text{i}/2+\text{Re}(\delta_{k20}))$ to second order, giving us
\begin{equation}\label{dk_20_app}
    \text{Re}(\delta_{k20}^{\pm})=-\frac{b_{k}(\text{i}/2)}{2a_{k}(\text{i}/2)}\left(1 \mp \sqrt{1-\frac{4a_{k}(\text{i}/2)c_{k}(\text{i}/2,\ups)}{b^2_{k}(\text{i}/2)}} \right),
\end{equation}
where $\text{Re}(\delta_{k20}^{+})$ and $\text{Re}(\delta_{k20}^{-})$ correspond to the decay rates of the slowly and quickly decaying wave respectively. We can further approximate $\text{Re}(\delta_{k20}^{+})$ near the most unstable wavenumber $k_{F20}$ by
\begin{equation}\label{df20_eq}
    \text{Re}(\delta^{+}_{k20}) \approx \text{Re}(\delta_{F20}^{+}) - D_{20} (k-k_{F20})^2
\end{equation}
where $\text{Re}(\delta_{F20}^{+})=\lim_{k\to k_{F20}}\text{Re}(\delta_{k20}^{+})$ and $D_{20}=-\tfrac{1}{2}\text{d}^2\text{Re}(\delta^{+}_{k20})/\text{d}k^2|_{k=k_{F20}}$ is the diffusion coefficient corresponding to this Faraday window. These approximations of $\text{Re}(\delta^{+}_{k20})$ from \eqref{dk_20_app} and \eqref{df20_eq} are shown in figure~\ref{fig: dk_plots}(a) as a blue dotted and a grey dashed curve respectively.

\begin{figure}
\centering
\includegraphics[width=\columnwidth]{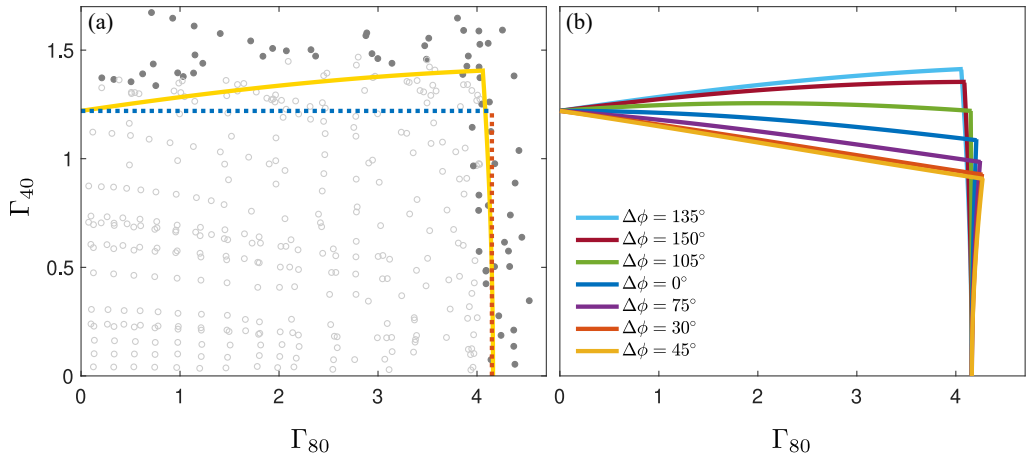}
\caption{Faraday threshold curves for two-frequency driving. (a) Comparison of the Faraday threshold curves for $\Delta\phi=130^{\circ}$ obtained using $21$ modes (solid yellow curves) and using the two-mode approximations (dotted curves) together with the experimental results (circles) of \citet{superwalker}. For the latter, empty circles indicate that flat liquid surfaces were observed while filled circles indicate that Faraday waves were observed. (b) Faraday thresholds for different values of the phase difference $\Delta\phi$ using a $21$-mode calculation.}
\label{fig: Far_ins}
\end{figure}

When $\text{Re}(\delta_k)> 0$ for any wavenumber $k$, growing Faraday waves are predicted. For two-frequency driving at $f$ and $f/2$, either $f/2$ Faraday waves or  $f/4$ Faraday waves can emerge depending on the relative strength of the acceleration amplitudes and the phase difference \citep{PhysRevLett.71.3287,superwalker}. The marginal stability curves representing the acceleration amplitudes at onset of Faraday waves, $\Gamma_{F80}$ and $\Gamma_{F40}$, can be found by setting $\text{Re}(\delta_k)=0$ when solving \eqref{inf_det}. From figure~\ref{fig: dk_plots}(a), we see two Faraday windows where $\text{Re}(\delta_k)$ can potentially cross zero corresponding to either the $f/2$ instability of frequency $40$\,Hz or the $f/4$ instability of frequency $20$\,Hz. Figure~\ref{fig: Far_ins}(a) shows the comparison of the numerically converged marginal stability curve obtained using a $21$-mode truncation (yellow solid curve) and the two-mode approximation for the $20$\,Hz (blue dashed curve) and $40$\,Hz (red dashed curve) Faraday waves, with the experiments of \citet{superwalker} (circles). Figure~\ref{fig: Far_ins}(b) shows the numerically converged marginal stability curves at different phase differences $\Delta\phi$. We note that changes in $\Delta\phi$ cause appreciable changes in the $20$\,Hz Faraday threshold.

\subsection{Amplitude and phase shift of damped Faraday waves}

In figures~\ref{fig: dk_plots}(d) and (e), the relative amplitudes of the Floquet modes are shown for the slowest decaying modes in the $20$\,Hz and $40$\,Hz Faraday windows respectively.  We now turn to calculating these amplitudes for our reduced-mode approximations and use these to obtain expressions for the wave profile generated by a single bounce of a droplet.

Using the two-mode approximation for the $40$\,Hz window, we can write the Floquet ansatz in \eqref{eq: flo_ana} as
\begin{equation}\label{3ma}
h_{\mathbf{k}}(\tau) \approx \left(h^{(-1)}_k\text{e}^{-\text{i}\tau}  + h^{(1)}_k\text{e}^{\text{i}\tau}\right) \text{e}^{\delta_{k}\tau},
\end{equation}
and the infinite dimensional linear system $\mathsfbi{A} \mathbf{h}=\boldsymbol{0}$ reduces to a $2\times 2$ matrix system $\mathsfbi{A}_2 \mathbf{h}_2=\boldsymbol{0}$. Since the determinant of the matrix $\mathsfbi{A}_2$ is zero, we obtain the amplitudes $\mathbf{h}_2$ from the null space vector, which gives $h^{(-1)}_k=c\,\xi^{\pm}_{40}$ with $\xi^{\pm}_{40}=\text{i}\sqrt{\alp/f_k(-\text{i}+\delta^{\pm}_{k40})}$ and $h^{(1)}_k=\overline{h^{(-1)}_k}$, where $c$ is a free parameter. 
 Substituting this solution into \eqref{3ma} and using $\text{Im}(\delta^{\pm}_{k40})=0$ in this window, we obtain the wave forms
\begin{equation*}
h^{\pm}_{\mathbf{k}40}(\tau) = c \left[\xi^{\pm}_{40}\text{e}^{-\text{i}\tau} + \overline{\xi}^{\pm}_{40}\text{e}^{\text{i}\tau}\right] \text{e}^{\text{Re}(\delta^{\pm}_{k40})\tau}.
\end{equation*}

Thus, the total wave field in this Faraday window can be represented as
\begin{equation}\label{totalwave40}
h_{\mathbf{k}40}(\tau) = \zeta_{40}^{+} \text{e}^{\text{Re}(\delta_{k40}^{+})\tau}\cos(\tau+\theta_{k40}^{+}) + \zeta_{40}^{-}  \text{e}^{\text{Re}(\delta_{k40}^{-})\tau}\cos(\tau+\theta_{k40}^{-}),
\end{equation}
where
\begin{equation}\label{40wave_par}
\theta_{k40}^{\pm}=\arctan\left(\frac{-\text{Im}(\xi_{40}^{\pm})}{\text{Re}(\xi_{40}^{\pm})} \right).
\end{equation}
and $\zeta^{\pm}_{40}=2c|\xi^{\pm}_{40}|$. These equations \eqref{40wave_par} and \eqref{totalwave40} are equivalent to equations $(2.47)$ and $(2.48)$ of \citet{tadrist_shim_gilet_schlagheck_2018}. Similar to Section 2.3 of \citet{tadrist_shim_gilet_schlagheck_2018}, we now continue by modelling the temporal profile of a droplet's impact by a delta function. The corresponding pressure and force exerted by the droplet on the liquid is then
\begin{equation*}
\Pi_{\mathbf{k}}(\tau)=(2k/\pi\Omega^2\rho)F_N(\tau)=\delta(\tau - \tau_i)v_{\mathbf{k}}.    
\end{equation*}
 By integrating the time domain version of equation \eqref{twofreqF_full} across the delta kick, we find that $v_{\mathbf{k}}$ corresponds to negative change of velocity of $h_\mathbf{k}$ following impact. If the surface is perfectly flat and at rest before the impact, the wave profile is axisymmetric i.e. $h_{\mathbf{k}}(\tau)=h_{k}(\tau)$. Using the initial conditions as $\tau \xrightarrow{} \tau_i$ that $h_{\mathbf{k}}=0$ and $\frac{\partial h_{\mathbf{k}}}{\partial \tau}=-v_k$ we get $\zeta_{40}^{\pm}=v_k \alpha_{40}^{\pm}$, where
\begin{equation*}
\alpha_{40}^{\pm}= \frac{-2\cos(\tau_i+\theta_{k40}^{\mp})\text{e}^{-\text{Re}(\delta_{k40}^{\pm})\tau_i}}{(\text{Re}(\delta_{k40}^{\pm})-\text{Re}(\delta_{k40}^{\mp}))(\cos(2\tau_i+\theta_{k40}^{\pm}+\theta_{k40}^{\mp})+\cos(\theta_{k40}^{\pm}-\theta_{k40}^{\mp}))-2\sin(\theta_{k40}^{\pm}-\theta_{k40}^{\mp})}.
\end{equation*}

Taking a similar approach, we can obtain an equation for the wave field by using the two-mode approximation in the $20$\,Hz Faraday window. The two-mode approximation of \eqref{eq: flo_ana} gives
\begin{equation*}
h^{\pm}_{\mathbf{k}20}(\tau) = (h^{(-1)}_k\text{e}^{-\text{i}\tau/2} + h^{(0)}_k \text{e}^{\text{i}\tau/2}) \text{e}^{\text{Re}(\delta^{\pm}_{k20})\tau}.
\end{equation*}

Solving for the null space of the matrix equation we get $h^{(-1)}_k=c\xi^{\pm}_{20}$ with $\xi^{\pm}_{20}=\text{i}\sqrt{\ups/f_k(-\text{i}/2+\text{Re}(\delta^{\pm}_{k20}))}$ 
and $h^{(0)}_k=\overline{h^{(-1)}_k}$, where $c$ is a free parameter.

For this we obtain the  amplitudes
\begin{equation*}
h^{\pm}_{\mathbf{k}20}(\tau) = c(\xi^{\pm}_{20}\text{e}^{-\text{i}\tau/2} + \overline{\xi}^{\pm}_{20}\text{e}^{\text{i}\tau/2}) \text{e}^{\text{Re}(\delta^{\pm}_{k20})\tau}.
\end{equation*}
Hence, we can express the total wave field for this Faraday window as
\begin{equation}\label{totalwave20}
h_{\mathbf{k}20}(\tau) = \zeta_{20}^{+}\text{e}^{\text{Re}(\delta_{k20}^{+})\tau}\cos(\tau/2 + \theta_{k20}^{+})+\zeta_{20}^{-}\text{e}^{\text{Re}(\delta_{k20}^{-})\tau}\cos(\tau/2 + \theta_{k20}^{-}),
\end{equation}
where
\begin{equation}\label{20wav_par}
\theta_{k20}^{\pm} = \arctan\left( \frac{-\text{Im}(\xi_{20}^{\pm})}{\text{Re}(\xi_{20}^{\pm})} \right).
\end{equation}
Using the same initial conditions as for $40$\,Hz waves we get $\zeta_{20}^{\pm}=v_k\alpha_{20}^{\pm}$, where
\begin{equation*}
\alpha_{20}^{\pm} =\frac{-2\cos(\tau_i/2+\theta_{k20}^{\mp})\text{e}^{-\text{Re}(\delta_{k20}^{\pm})\tau_i}}{(\text{Re}(\delta_{k20}^{\pm})-\text{Re}(\delta_{k20}^{\mp}))(\cos(\tau_i+\theta_{k20}^{\pm}+\theta_{k20}^{\mp})+\cos(\theta_{k20}^{\pm}-\theta_{k20}^{\mp}))-2\sin(\theta_{k20}^{\pm}-\theta_{k20}^{\mp})}. 
\end{equation*}

\subsection{Wave field of a superwalker}

For late times after the impact, $\tau\gg\tau_i$, and when the acceleration amplitudes are close to their respective Faraday thresholds, $\Gamma_{80}\lesssim\Gamma_{F80}$ and $\Gamma_{40}\lesssim\Gamma_{F40}$, the wave field is dominated by the slowly decaying Faraday waves from both the $40$\,Hz and $20$\,Hz modes. Hence, we can approximate the final wave field generated by the impact of the droplet as
\begin{equation*}
h_{\mathbf{k}}(\tau)= \alpha_{40}^{+}v_k \text{e}^{\text{Re}(\delta_{k40}^{+})\tau} \cos(\tau+\theta_{k40}^{+})+\alpha_{20}^{+}v_k\text{e}^{\text{Re}(\delta_{k20}^{+})\tau}\cos(\tau/2 + \theta_{k20}^{+}).
\end{equation*}

Transforming back to the spatial domain with an inverse Fourier transform yields
\begin{equation*}
h(x,y,\tau)=\frac{1}{2\pi}\int_0^\infty \int_0^\infty h_{\mathbf{k}}(\tau) \text{exp}[{\text{i}(k_x x +k_y y)}] \,\text{d}k_x \,\text{d}k_y   \, .
\end{equation*}
Since the wave profile is radially symmetric, the above inverse Fourier transform reduces to an inverse Hankel transform,
\begin{equation*}
h(x,y,\tau)=\int_0^\infty h_{\mathbf{k}}(\tau) \text{J}_0(kr)k\text{d}k.   
\end{equation*}
Hence the wave profile in the real space is given by
\begin{align*}
h(x,y,\tau)&=\int_0^\infty B_{k40}^{+}v_k \text{e}^{\text{Re}(\delta_{k40}^{+})(\tau-\tau_i)} \cos(\tau+\theta_{k40}^{+})\text{J}_0(kr)k\text{d}k\\
&+ \int_0^\infty B_{k20}^{+}v_k\text{e}^{\text{Re}(\delta_{k20}^{+})(\tau-\tau_i)}\cos(\tau/2 + \theta_{k20}^{+})  \text{J}_0(kr)k\text{d}k , 
\end{align*}
where $B_{k40}^{+}=\alpha_{40}^{+}\text{e}^{\text{Re}(\delta_{k40}^{+})\tau_i}$ and $B_{k20}^{+}=\alpha_{20}^{+}\text{e}^{\text{Re}(\delta_{k20}^{+})\tau_i}$. 
Using the second order expansion for $\text{Re}(\delta^{+}_{k40})$ and $\text{Re}(\delta^{+}_{k20})$ in \eqref{df40_eq} and \eqref{df20_eq}, we get the following approximation to the above integral in the limit $\tau\xrightarrow{}\infty$ \cite[for details see Appendix C of][]{tadrist_shim_gilet_schlagheck_2018}
\begin{align*}
h(x,y,\tau)&= \Tilde{A}^{(0)}_{40}(\tau_i)\frac{\cos(\tau+\theta_{F40}^{+})}{\sqrt{\tau-\tau_i}} \text{J}_0(k_{F40}|\mathbf{x}-\mathbf{x}_i|)\,\exp{\left[ -\frac{\tau-\tau_i}{2\pi \text{Me}_{40}} - \frac{|\mathbf{x}-\mathbf{x}_i|^2}{4 D_{40} (\tau-\tau_i)} \right]}\notag\\
&+ \Tilde{A}^{(0)}_{20}(\tau_i)\frac{\cos(\tau/2 + \theta_{F20}^{+})}{\sqrt{\tau-\tau_i}} \text{J}_0(k_{F20}|\mathbf{x}-\mathbf{x}_i|)\,\exp{\left[ -\frac{\tau-\tau_i}{2\pi\text{Me}_{20}} - \frac{|\mathbf{x}-\mathbf{x}_i|^2}{4 D_{20} (\tau-\tau_i)} \right]},
\label{hxyt}
\end{align*}
where $\mathbf{x}_i$ is the location of the impact and the memory parameters $\text{Me}_{40}$ and $\text{Me}_{20}$ are given by $\text{Me}_{40}=-1/2\pi\text{Re}(\delta_{F40}^{+})$ and $\text{Me}_{20}=-1/2\pi\text{Re}(\delta_{F20}^{+})$. Furthermore,
\begin{equation*}
    \Tilde{A}^{(0)}_{40}=k_{F40}\sqrt{\frac{\pi}{D_{40}}}v_{k}B_{F40}^{+}(\tau_i)\,\,\text{and}\,\,\Tilde{A}^{(0)}_{20}=k_{F20}\sqrt{\frac{\pi}{D_{20}}}v_{k}B_{F20}^{+}(\tau_i).
\end{equation*}
To include a finite contact time, we follow the suggestion in \citet{tadrist_shim_gilet_schlagheck_2018} of using Duhamel's principle and the approach used in Appendix A.4 of \citet{Molacek2013DropsTheory}, and integrate the impulse response with a time varying impact signal $\Pi_{\mathbf{k}}(\tau)$. This results in replacing the amplitude coefficients $\Tilde{A}^{(0)}_{40}$ and $\Tilde{A}^{(0)}_{20}$ by 
\begin{equation*}
    \Tilde{A}_{40}=k_{F40}\sqrt{\frac{\pi}{D_{40}}}\int_{\tau_n^i}^{\tau_n^c}B^{+}_{F40}(\tau') \Pi_{\mathbf{k}}(\tau') \,\text{d}\tau',
\end{equation*}
\begin{equation*}
    \Tilde{A}_{20}=k_{F20}\sqrt{\frac{\pi}{D_{20}}}\int_{\tau_n^i}^{\tau_n^c}B^{+}_{F20}(\tau') \Pi_{\mathbf{k}}(\tau')\,\text{d}\tau'.
\end{equation*}
We change the dimensionless time $\tau$ back to dimensional time $t$ and replace $\Pi_{\mathbf{k}}(t)$ by $(2k/\pi \Omega^2 \rho)F_N(t)$ using \eqref{eq: pi_force}. We also replace the initial contact time $t_i$ and location of contact $\mathbf{x}_i$ by their weighted average values $t_n$ and $\mathbf{x}_n$ as given in \eqref{wa_pos}, and replace the dimensionless amplitudes $\Tilde{A}_{40}$ and $\Tilde{A}_{20}$ by $A_{40}=\sqrt{2/\Omega}\Tilde{A}_{40}$ and $A_{20}=\sqrt{2/\Omega}\Tilde{A}_{20}$ which gives \eqref{amp_sw} and results in the wave field equation \eqref{sw_wave}.

\begin{figure}
\centering
\includegraphics[width=\columnwidth]{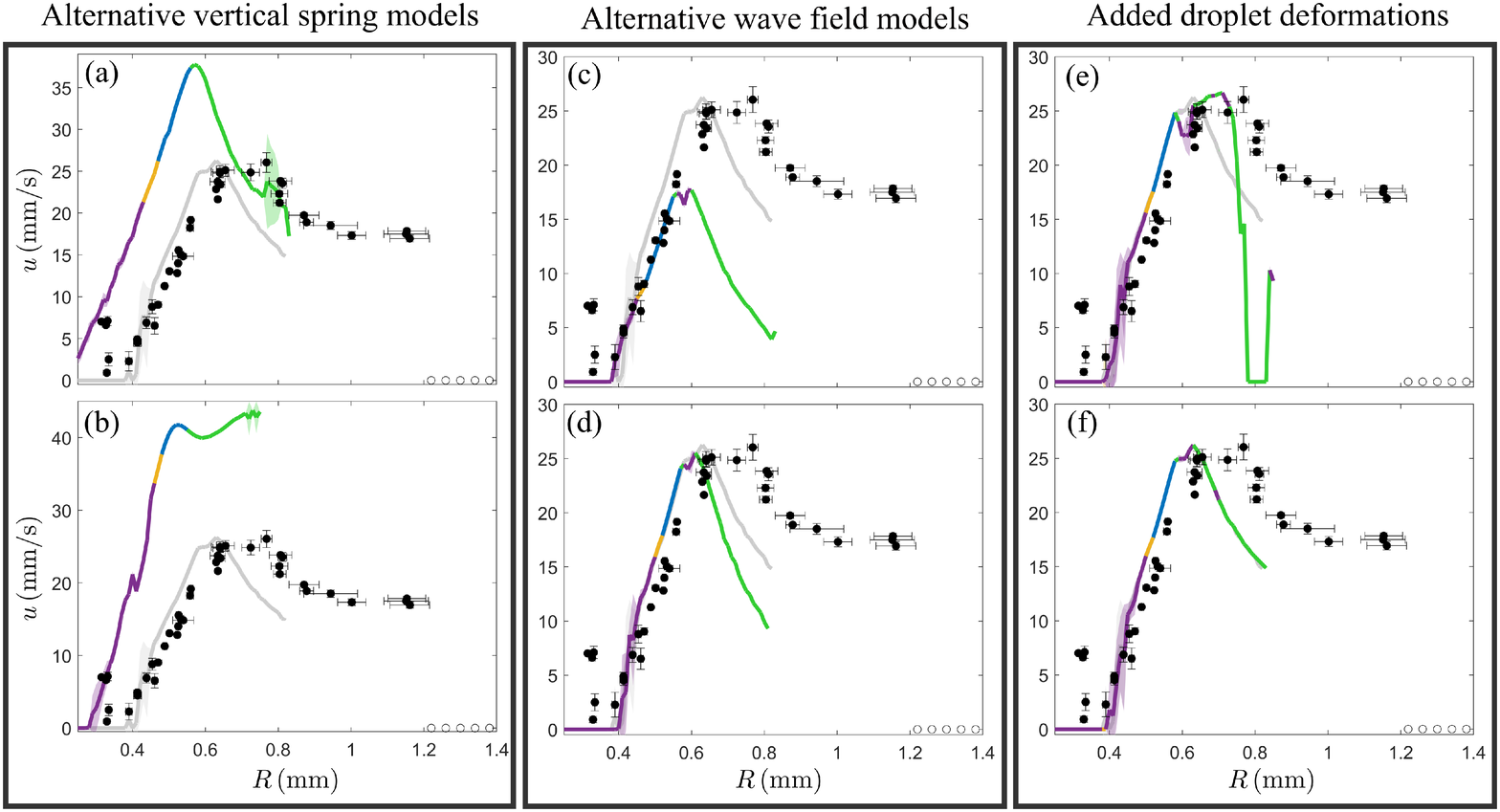}
\caption{Comparison of the speed-size characteristics of different models at $\Gamma_{80}=3.8$, $\Gamma_{40}=0.6$ and $\Delta\phi=130^{\circ}$. In each panel, the black circles are experimental results of \citet{superwalker}, grey curves are the results using the model presented in this paper and the coloured curves are results from different models stated below with the colour indicating bouncing mode using the same conventions as in figure~\ref{fig: SpeedSize}. Termination of the solid curves indicate coalescence. Results of using two alternative vertical spring models, a simple linear spring model and the logarithmic spring model of \citet{molacek_bush_2013}, are shown in (a) and (b), respectively. Results obtained using a wave field from the model of \citet{Molacek2013DropsTheory} and \citet{tadrist_shim_gilet_schlagheck_2018} are shown in (c) and (d), respectively. Results obtained by adding droplet deformation based on \citet{deformationdroplet} and \citet{Gilet_2008} are shown in (e) and (f), respectively. For the grey curves and the coloured curves in all panels except (b), the linear spring model was used for the vertical dynamics with the parameters $K$ defined according to \eqref{kb_linear} and a fixed $B=0.60$.}
\label{fig: OM}
\end{figure}

\section{Comparison of different droplet models}\label{compare models}

To test the robustness of the superwalking behaviour, we explored superwalkers using alternative models for the vertical dynamics, the wave field generated, and adding droplet deformations to the model presented in \S\:\ref{NM}. Comparison of these models with the model presented in this paper and the experimental results of \citet{superwalker} for a typical speed-size curve of superwalkers is presented in figure~\ref{fig: OM}.

Apart from the linear spring model used in this work, two alternative spring models for the vertical dynamics of a bouncing droplet were presented by \citet{molacek_bush_2013}: (i) a simple linear spring model that does not restrict the normal force to be positive \emph{i.e.,} without the maximum condition in \eqref{springdamp}, and (ii) a logarithmic spring model, which can be implemented by replacing \eqref{eq: vertical} with
\begin{equation*}
    \left( 1+\frac{C_3}{\ln^2{\big{|} \frac{C_1R}{\bar{z}_d}\big{|}}} \right)m\ddot{z}_d+\frac{4}{3}\frac{\pi \nu \rho R C_2}{\ln^2{\big{|}\frac{C_1R}{\bar{z}_d}\big{|}}}\dot{\bar{z}}_d+\frac{2\pi\sigma}{\ln{\big{|}\frac{C_1R}{\bar{z}_d}\big{|}}}\bar{z}_d=-m[\text{g}+\gamma(t)]
\end{equation*}
when the droplet is in contact with the bath, and using $m\ddot{z}_d=-m[\text{g}+\gamma(t)]$ when the droplet is in the air. We fixed the parameter values to $C_1=2$, $C_2=12.5$ and $C_3=1.4$ which are typical values used for walkers \citep{molacek_bush_2013}. Coupling these vertical dynamics models with the wave field and the horizontal dynamics described in \S\:\ref{NM}, we obtain the speed-size curves presented in figures~\ref{fig: OM}(a) and (b). 

Using the wave field of a walker from the \citet{Molacek2013DropsTheory} model presented in \eqref{molacek wave} and the \citet{tadrist_shim_gilet_schlagheck_2018} model presented in \eqref{tadrist_wave} in place of the superwalker wave field that was used in this work, we obtain the speed-size curves shown in figures~\ref{fig: OM}(c) and (d). These curves also show good match with the experiments on the ascending branch. We note that for a droplet in a \otoh{} bouncing mode, the subsequent bounce would occurs one Faraday period after the initial impact. At this time, there is approximately a 10\% difference in the amplitudes between the three models, and a slightly greater difference in the gradients (see figure~\ref{fig: wavefield}). This would suggest a comparable difference in the walking speeds. However, although in figure~\ref{fig: OM}(c), the peak of the speed-size curve from the wave model of \citet{Molacek2013DropsTheory} only goes up to approximately $17$\,mm/s for the present choice of $K$ and $B$ values, we obtain a better fit to the experimental results by alternate choices of parameters $K$ and $B$. Hence by tuning the $K$ and $B$ values and using the wave model of \citet{Molacek2013DropsTheory}, we can obtain good fit to the experimental data which is comparable to the fit obtained from the superwalker wave model. Speed-size curve from the wave model of \citet{tadrist_shim_gilet_schlagheck_2018} is identical to the curve from the superwalker wave model on the ascending branch. On the descending branch, we see that lower speeds are obtained from the \citet{tadrist_shim_gilet_schlagheck_2018} model compared to the superwalker wave field. This shows that the added $20$\,Hz waves seems to slightly speed up larger droplets on the descending branch in \otol{} bouncing mode. 

Finally, to account for droplet deformations, we couple the droplet deformation models of \citet{deformationdroplet} and \citet{Gilet_2008} to the theoretical model presented in \S\:\ref{NM}. The additional droplet deformation equation for the model of \citet{deformationdroplet} is
\begin{equation*}
    m\ddot{R}_v+c_d\dot{R}_v+m\omega^2(R_v-R)=-F_N(t),
\end{equation*}
where $R_v$ is the vertical radius of the droplet, $c_d=3.8 m \nu/R^2$ is the effective damping coefficient of the droplet deformation, and $\omega=\sqrt{N_{\omega}\sigma/\rho R^3}$ is the droplet's natural frequency with $N_{\omega}=5.84$. The model of \citet{Gilet_2008} after some algebra also reduces to an effectively similar equation for droplet deformations and is given by
\begin{equation*}
    c_3m\ddot{R}_v+\frac{c_5 m \nu}{R^2}\dot{R}_v+c_4\sigma(R_v-R)=-c_6 F_N(t),
\end{equation*}
where the parameters $c_3=0.1$, $c_4=10$, $c_5=3.3$ and $c_6=1$. While implementing both of these models, the criteria for contact changes from $\bar{z}_d\leq0$ to $\bar{z}_d+R-R_v\leq0$. Coupling these droplet deformation models to the theoretical model in \S\:\ref{NM} results in the speed-size curves shown in figures~\ref{fig: OM}(e) and (f). We see that the model of \citet{Gilet_2008} seems to have an insignificant effect on the speed-size characteristics with the curves completely overlapping each other. The model of \citet{deformationdroplet} increases the walking speed of droplets in a small neighbourhood around $R=0.7$\,mm but the model is still unable to capture the large superwalkers.

\section{Determination of parameters $K$ and $B$}\label{KBR}

\begin{figure}
\centering
\includegraphics[width=\columnwidth]{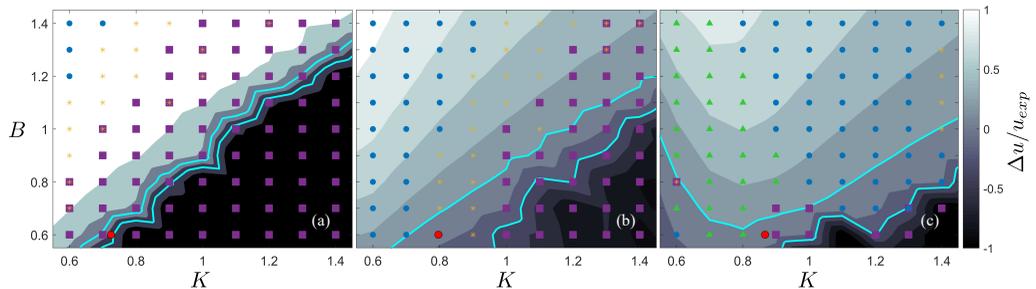}
\caption{Determination of the parameter values $K$ and $B$. Bouncing modes (markers) and relative difference between the numerical and the experimental values of the walking speed $\Delta u/u_{\text{exp}}=(u-u_{\text{exp}})/u_{\text{exp}}$ (contours) in the $(K,B)$ parameter space for three different droplet radii (a) $R=0.5$\,mm, (b)$R=0.6$\,mm and (c)$R=0.7$\,mm at $\Gamma_{80}=3.8$, $\Gamma_{40}=1$ and $\Delta\phi=130^{\circ}$. In all the three panels, blue circles $\bullet$ are \otoh{}, green triangles {\protect \scalebox{0.8}{$\blacktriangle$}} are \otol{}, yellow asterisks $*$ are (2,4,2) and purple squares {\protect \scalebox{0.5}{$\blacksquare$}} represent chaotic or 
other higher periodicity bouncing modes. The cyan solid lines represent the boundaries of the region inside which $|\Delta u/u_{\text{exp}}|<20\%$ and the red circle corresponds to our chosen $K$ according to \eqref{kb_linear} and a fixed $B=0.60$.}
\label{fig: KB}
\end{figure}

The theoretical model for simulating superwalkers presented in \S\:\ref{NM} has three free parameters that are not known for superwalkers: (i) the dimensionless spring constant $K$ (ii) the dimensionless damping coefficient $B$ and (iii) the contact drag coefficient $C$. In our study we fixed $C=0.17$, a typical value that is used for walkers \citep{Molacek2013DropsTheory}. To determine values of $K$ and $B$, we simulated superwalkers in the $(K,B)$ parameter space and selected values that provide a good fit to the experimental results of \citet{superwalker}. We found that using a constant values of $K=0.70$ and $B=0.60$ provided a reasonably good fit for small- to moderate-sized superwalkers on the ascending branch of the speed-size curves presented in figure~\ref{fig: SpeedSize}, but failed for the  largest superwalkers on the ascending branch for $\Gamma_{40}=1$. By allowing $K$ to vary linearly with the droplet radius $R$ while keeping $B$ fixed to $0.60$, we were able to obtain a better fit on the ascending branch for the results presented in figure~\ref{fig: SpeedSize}. To arrive at this linear relationship, we simulated superwalkers for a fixed $\Gamma_{80}=3.8$, $\Delta\phi=130^{\circ}$ and four different values of $\Gamma_{40}=0$, $0.3$, $0.6$ and $1$. Droplet size that corresponds to the ascending branch in figure~\ref{fig: SpeedSize} were simulated. Typical graphs for the droplet speed and bouncing modes are shown in figure~\ref{fig: KB}. For each droplet size and $\Gamma_{40}=0.6$ and $1$, the region of the $(K,B)$ parameter space where the relative difference between the speed of simulated superwalker and the corresponding experimental value of $\Delta u/u_{\text{exp}}=(u-u_{\text{exp}})/u_{\text{exp}}$ is within $20\%$ was determined and then a value of $K$ was selected from that region that matched with the experimentally observed bouncing mode. A linear best fit through all such $K$ values for different sized droplets results in one generic linear relationship given in \eqref{kb_linear}.

\bibliographystyle{jfm}

\bibliography{emergence_superwalkers}

\end{document}